\documentclass[10pt,aps,showpacs,nofootinbib,prd,aps,epsf,floats,
               amsmath,amssymb,amsfonts,axodraw,floatfix,graphicx,twocolumn]{revtex4-1}

\usepackage{amsmath, amssymb}
\usepackage{float}
\usepackage{axodraw}
\usepackage{multirow}
\usepackage{paralist}
\newcommand{\mathsym}[1]{{}}

\usepackage{graphicx}
\usepackage{amsmath}
\usepackage{amssymb}
\usepackage{amsmath}
\usepackage{multirow}
\usepackage{paralist}
\usepackage{slashed}
\usepackage{amsfonts}
\usepackage{hyperref} 
\usepackage{graphicx}
\usepackage{amsmath}
\usepackage{amssymb}
\usepackage{mathrsfs}
\newsavebox{\PSLASH}
 \sbox{\PSLASH}{$p$\hspace{-1.8mm}/}
 
\renewcommand{\theequation}{\thesection.\arabic{equation}}
\newcounter{saveeqn}
\newcommand{\add}{\addtocounter{equation}{1}}
\newcommand{\alphaeqn}{\setcounter{saveeqn}{\value{equation}}%
\setcounter{equation}{0}%
\renewcommand{\theequation}{\mbox{\thesection.\arabic{saveeqn}{\alpha{equation}}}}}
\newcommand{\reseteqn}{\setcounter{equation}{\value{saveeqn}}%
\renewcommand{\theequation}{\thesection.\arabic{equation}}}

 \newsavebox{\notrightarrow}
 \sbox{\notrightarrow}{$\to$\hspace{-4mm}/}
 
 \newsavebox{\PARTIALSLASH}
 \sbox{\PARTIALSLASH}{$\partial$\hspace{-1.6mm}/}
 
 \newsavebox{\ASLASH}
 \sbox{\ASLASH}{$A$\hspace{-2.1mm}/}
 
 \newsavebox{\KSLASH}
 \sbox{\KSLASH}{$k$\hspace{-1.8mm}/}
 
 \newsavebox{\LSLASH}
 \sbox{\LSLASH}{$\ell$\hspace{-1.8mm}/}
 
 \newsavebox{\QSLASH}
 \sbox{\QSLASH}{$q$\hspace{-1.8mm}/}
 
 \newsavebox{\DSLASH}
 \sbox{\DSLASH}{$D$\hspace{-2.2mm}/}
 
 \newsavebox{\DbfSLASH}
 \sbox{\DbfSLASH}{${\mathbf D}$\hspace{-2.8mm}/}
 
 \newsavebox{\DELVECRIGHT}
 \sbox{\DELVECRIGHT}{$\stackrel{\rightarrow}{\partial}$}
 
 \newcommand{\blue}{\IfColor{\textCadetBlue}{}}
\newcommand{\black}{\IfColor{\textBlack}{}}
\newcommand{\red}{\IfColor{\textRed}{}}
\newcommand{\green}{\IfColor{\textOliveGreen}{}}
\newcommand{\lil}{\IfColor{\textRedViolet}{}}

\newcommand\abs[1]{\left|#1\right|}

\newcommand{\bs}{\boldsymbol}

\makeatother
\usepackage[T1]{fontenc}
\usepackage[latin9]{inputenc}
\usepackage{amsmath}
\usepackage{amssymb}
\usepackage{graphicx}
\usepackage{dcolumn}
\usepackage{verbatim}
\usepackage{orcidlink}

\begin{document}
\title{Weak Bose-Einstein condensation in a rigidly rotating magnetized charged Bose gas}
\author{E. Siri $^{a,b}$\,~\orcidlink{0009-0008-1021-3348}~~}\email{e.siri@physics.sharif.ir}\author{N. Sadooghi $^{a}$\,~\orcidlink{0000-0001-5031-9675}~~}\email{Corresponding author: sadooghi@physics.sharif.ir}
\affiliation{$^{a}$Department of Physics, Sharif University of Technology,
P.O. Box 11155-9161, Tehran, Iran}
\affiliation{$^{b}$Research Center for High Energy Physics, Department of Physics, Sharif University of Technology, P.O. Box 11155-9161, Tehran, Iran}
\begin{abstract}
We investigate the weak Bose-Einstein condensation (BEC) scenario of a noninteracting charged Bose gas simultaneously subjected to a strong magnetic field and rigid rotation. Using standard methods of finite-temperature quantum field theory and the generalized Fock-Schwinger formalism, we derive the corresponding thermodynamic potential in the nonrelativistic and lowest Landau level approximations. An appropriate modification of the effective chemical potential yields a consistent thermodynamic description and naturally introduces a magnetorotational fugacity. Within the high-temperature approximation, rigid rotation enters the thermodynamics solely through the Tolman-Ehrenfest local temperature. We demonstrate that rigid rotation does not qualitatively modify the weak BEC scenario induced by Landau quantization. The magnetorotational fugacity remains below unity throughout the phenomenologically relevant temperature range, while the continuous evolution of the ground state population and the absence of a singularity in the specific heat provide complementary signatures of the persistence of weak BEC. We further study the thermodynamic properties of the system under conditions relevant to quark-gluon plasma and neutron-star matter. We show that rotational effects are much more pronounced in the former. Our analysis reveals a new magnetic response to rigid rotation: while magnetic fields enhance diamagnetism, rotation drives it toward paramagnetism. This behavior reflects a competition between magnetic quantization and rotational orbital motion, emphasizing the role of rotation in shaping the magnetic response of bosonic matter.
\end{abstract}
	\maketitle
\section{introduction}\label{sec1}
Bose-Einstein condensation (BEC) of charged scalar particles under extreme conditions has attracted considerable interest in relativistic many-body physics and astrophysics. In the presence of a strong magnetic field, charged bosons experience Landau quantization, which reduces the effective dimensionality of the system \cite{gusynin1996} and qualitatively modifies the condensation mechanism \cite{rojas1996}. In the absence of a magnetic field, BEC occurs at a finite critical temperature $T_c$. However, when a magnetic field is applied, the sharp phase transition typically associated with BEC disappears. Instead, the system follows a weak condensation scenario, known as a diffuse phase transition \cite{rojas1996,rojas2025}. In this scenario, the ground state occupation increases gradually as the temperature decreases, without a true thermodynamic transition. This behavior has been established for magnetized charged bosons and has important implications for astrophysical environments, such as neutron stars \cite{rojas2025}. In these environments, pion condensation may affect cooling rates and the equations of state \cite{hashimoto2021}.
\par
Studies of charged bosons with quartic self-interactions in a magnetic field have shown that the system exhibits magnetic catalysis (MC), in which the magnetic field favors the formation of the condensate \cite{ayala2012,ayala2016}. When plasma screening effects are properly included, a well-defined $T_c$ appears and increases with the magnetic field strength \cite{ayala2012}.
For noninteracting charged scalar bosons in the weak condensation scenario, the situation is qualitatively different. Under neutron star conditions, the condensation occurs in a two-step process \cite{rojas2025}. First, particles accumulate in the lowest Landau level (LLL), which is then followed by a gradual concentration of particles near the true ground state. During this process, the system experiences a diffuse phase transition, without a sharp $T_c$. Consequently, no peaks are observed in the heat capacity. Additionally, the magnetization of the system shifts from a diamagnetic to a paramagnetic state at a certain magnetic field threshold. This crossover is accompanied by significant changes in heat capacity, magnetization, and pressure once antiparticles and higher Landau levels (HLL) contributions are included.
\par
The influence of rotation on Bose gases has attracted considerable attention in recent years. In the absence of an external magnetic field, rigid rotation can induce thermodynamic instabilities in interacting bosonic systems. For a $\lambda\varphi^{4}$ theory, these instabilities may manifest through negative moment of inertia and heat capacity in certain interaction regimes. Rigid rotation also significantly modifies the thermodynamics of a noninteracting Bose gas by suppressing BEC, reducing the critical temperature, and inducing discontinuities in the specific heat within the nonrelativistic regime \cite{siri2024-2}. The interplay between rigid rotation and spontaneous symmetry breaking has likewise revealed several novel features. In an interacting bosonic gas, rotation modifies the $\mathrm{U}(1)$ symmetry breaking transition, leading to a critical temperature that scales as $T_c \propto  \Omega^{1/3}$ \cite{siri2025}. Although the Goldstone theorem remains valid once rotation-dependent one-loop thermal corrections to the masses are included, nonperturbative effects qualitatively alter the nature of the transition, changing it from second order to a crossover. These studies demonstrate that rigid rotation can profoundly affect both the thermodynamic and critical properties of bosonic matter.
\par
Fermionic systems have been studied in the presence of magnetic fields \cite{gusynin1996,shovkovy2007,fayazbakhsh2011,schmitt2011,bali2012,fayazbakhsh2012,bruckmann2013,fayazbakhsh2013,ferrer2015,fayazbakhsh2014,hattori2017,adhikari2026}, rigid rotation \cite{yamamoto2013,ambrus2014,ambrus2016,jiang2016,ebihara2017,chernodub2017a,chernodub2017b,wang2019,chernodub2021,braguta2024a,pradhan2024,braguta2024b,morales2025,abedlou2025,singha2025}, and their simultaneous combined effects \cite{mameda2015,fukushima2015,zahed2018,sadooghi2021,hooman2023,fukushima2025}. In contrast, for bosons, despite separate studies of BEC in the presence of magnetic fields and rotation, a question remains. Can rigid rotation modify the weak BEC scenario or does it only change the thermodynamic properties of the system? This question is particularly pressing for realistic systems where both effects are present simultaneously. Noncentral relativistic heavy-ion collisions (HICs) generate magnetic fields of order $10^{18} - 10 ^{19}$ G \cite{skokov2009,huang2012} alongside angular momenta of $10^{3} - 10^{4} \hbar$ \cite{liao2016,star2017,becattini2020}. Similarly, neutron stars and magnetars host strong magnetic fields ($\gtrsim10^{15}$ G) \cite{duncan1992,peng2007} combined with rapid rotation \cite{beloborodov2017}. Understanding the interplay between these two mechanisms is essential for determining whether pion condensation in such environments acquires qualitatively new feature or remain within the weak condensation paradigm. In the present work, we aim to address this question.
\par
In the context of magnetic fields and rigid rotation, Liu and Zahed proposed the possibility of charged pion condensation at finite temperatures while preserving charge neutrality \cite{zahed2016}. However, later analyses  indicated that such condensation does not occur in the noninteracting limit due to dimensional reduction, in accordance with the Coleman-Mermin-Wagner-Hohenberg theorem \cite{lianyi2025}. Subsequent studies explored the effects of parallel magnetic fields and rotation, suggesting that charged bosons could form a Bose-Einstein condensate and lead to the formation of supergiant quantum vortices, which have significant implications for quark-gluon plasma (QGP) in HICs \cite{lianyi2022,guo2024,voskresensky2024-1}. Additionally,
studies of the stability of pion condensation showed that rapid rotation can significantly affect it \cite{voskresensky2024-2}. Despite predictions from models like the Nambu-Jona--Lasinio (NJL) suggesting uniform condensation under specific conditions, more detailed analyses indicate that complex inhomogeneous configurations, such as vortex lattices, are more likely at high rotation rates \cite{cao2019,chen2024}.
A recent study explored the transport properties of a rotating charged pion gas in a strong magnetic field \cite{vinod2026}. It was shown that the interplay between rotation and magnetic field leads to rich behavior in thermoelectric responses.
\par
Whether rigid rotation can modify the weak BEC scenario or merely alters the thermodynamic properties of the system remains an open question. The present work addresses this question by developing a finite-temperature field theory framework for charged scalar bosons in the presence of a magnetic field and rigid rotation.
We explore two different phenomenologically motivated parameter regimes corresponding to quark-gluon plasma, characterized by relatively large magnetic fields and angular velocities, and neutron star matter,
where both quantities are considerably smaller.
Throughout this work, we do not impose global charge neutrality, since our objective is not to construct a complete microscopic description of these systems, but rather to isolate the combined effects of magnetic fields and rigid rotation on the weak condensation mechanism.
Using the generalized Fock-Schwinger formalism, we derive the propagator of free charged bosons at finite temperature and compute the corresponding thermodynamic potential. We focus on the nonrelativistic limit and LLL approximation, modifying the effective chemical potential to obtain a thermodynamically consistent thermodynamic potential. This construction naturally leads to the definition of the magnetorotational fugacity $z_{B,\Omega}$. Within the high-temperature approximation, rotation enters the thermodynamic description solely through the Tolman-Ehrenfest local temperature \cite{tolman1930,ehrenfest1930}.
\par
Our analysis shows that the weak BEC scenario persists even in the presence of rigid rotation. In contrast to previous studies that imposed charge neutrality and included both particle and antiparticle contributions  \cite{zahed2016, lianyi2025}, our framework isolates the particle sector and employs a high-temperature expansion to determine $z_{B,\Omega}$. We find that rigid rotation does not restore a true Bose-Einstein phase transition with a finite critical temperature. Instead it leads to significant quantitative modifications of the thermodynamic properties, including pressure, heat capacity, magnetization, and susceptibilities. The rotational effects are most pronounced in the QGP regime with much larger angular velocities. Furthermore, our results reveal a new magnetic response to rigid rotation: whereas magnetic fields enhance diamagnetism, rotation drives it toward paramagnetism.
\par
The organization of the paper is as follows: In Sec. \ref{sec2A}, we introduce the theoretical framework and derive the charged scalar propagator in a rotating medium with a background magnetic field, utilizing the generalized Fock-Schwinger method. In Sec. \ref{sec2B}, we determine the thermodynamic potential with a focus on particle contributions in the nonrelativistic limit for the lowest and higher Landau levels. In Sec. \ref{sec3}, we analyze the weak BEC scenario in the LLL approximation. Section IV presents thermodynamic relations in the corotating frame. Numerical results for QGP and neutron star regimes are given in Sec. \ref{sec5}. We conclude in Sec. \ref{sec6} with a summary of our findings and their physical implications. Appendix \ref{appA} collects useful thermodynamic relations.
\section{Theoretical framework}\label{sec2}
\setcounter{equation}{0}
\subsection{The model}\label{sec2A}
The main purpose of this paper is to investigate the effect of rigid rotation on BEC in the presence of a constant magnetic field. To describe the magnetized bosons, we use spin-zero complex Klein-Gordon (KG) fields, $\varphi$, whose dynamics is described by the Lagrangian density
\begin{eqnarray}\label{A1}
\mathscr{L}=\left( D_{\mu}\varphi \right)^{*}D^{\mu}\varphi-m^2\varphi^{*}\varphi,
\end{eqnarray}
where $m$ is the rest mass. We introduce the rigid rotation by the metric,
\begin{eqnarray}\label{A2}
	{{g}_{\mu \nu }}=\left( \begin{matrix}
		1-r^{2}{{\Omega }^{2}} & y\Omega  & -x\Omega  & 0  \\
		y\Omega  & -1 & 0 & 0  \\
		-x\Omega  & 0 & -1 & 0  \\
		0 & 0 & 0 & -1  \\
	\end{matrix} \right) ,
\end{eqnarray}
describing a rotation around the $z$ axis at a constant angular velocity $\Omega$. In \eqref{A2}, $x$ and $y$ are components of Cartesian coordinates $x^{\mu}=\left(t,x,y,z\right)$ and $r^{2}\equiv x^{2}+y^{2}$. In this paper, Greek indices $\alpha,\beta\in\{t,x,y,z\}$ refer to coordinates in the corotating frame, while Latin indices $a,b\in\{0,1,2,3\}$ denote coordinates in the laboratory frame. With this in mind, we introduce a constant background magnetic field in the third direction by fixing the gauge field $A^{a}$ in a symmetric gauge as $A^{a}=\left(0,-B_{0}y/2,B_{0}x/2,0\right)$ with $B_{0}>0$. This leads to $B^{a}=\left(0,0,0,B_{0}\right)$ in the laboratory frame. By utilizing $A^{\mu}=e^{\mu}_{~a}A^{a}$, it is possible to determine the gauge field in the corotating frame. Using the vierbeins $e^{\mu}_{~a}$, whose nonzero components are given by
\begin{eqnarray}\label{A3}
e^{t}_{0}=e^{x}_{1}=e^{y}_{2}=e^{z}_{3}=1, \quad e^{x}_{0}=y\Omega, \quad e^{y}_{0}=-x\Omega, \nonumber\\
\end{eqnarray}
we arrive at the gauge potential in corotating frame
\begin{eqnarray}\label{A4}
	A^{\mu}=\left(-B_{0}\Omega r^2/2,-B_{0}y/2,B_{0}x/2,0\right).
\end{eqnarray}
It is important to note that this gauge field implies, as expected, not only a constant magnetic field $\bs{B}=B_{0}\bs{e}_{z}$ but also a radial electric field $\bs{E}=B_{0}\Omega r\bs{e}_{r}$ in the corotating frame. Here, $\bs{e}_{r}$ is the unit vector in the radial direction within the cylindrical coordinate system defined by $x^{\mu}=\left(t,r\cos\theta, r\sin\theta,z\right)$. Here, $r$ represents the radius, $\theta$ the azimuthal angle, and $z$ the height of the cylinder.
\par
Plugging the covariant derivative $D_{\mu}\equiv \partial_{\mu}+ieA_{\mu}$ into \eqref{A1}, we obtain
\begin{eqnarray}\label{A5}
\mathscr{L}=|\widetilde{\partial}_t\varphi|^2-\left| D_i\varphi \right|^2-m^2\left| \varphi \right|^2,
\end{eqnarray}
with $\widetilde{\partial}_{t}\equiv \partial_t-i\Omega L_z$ and the $z$-component of the angular-momentum  $L_{z}\equiv i\left(y\partial_{x}-x\partial_{y}\right)$ or equivalently $L_{z}\equiv -i\partial_{\theta}$. The Euler-Lagrange equation of motion of a magnetized
KG field under rotation is thus given by
\begin{eqnarray}\label{A6}
\mathfrak{D}(x,\partial_{x})\varphi(x)=0,
\end{eqnarray}
with the differential operator
\begin{eqnarray}\label{A7}
\hspace{-0.5cm}\mathfrak{D}(x,\partial_x)\equiv \widetilde{\partial}_t^2-\bs{\nabla}^{2}-eB_0L_z+\frac{e^2B_{0}^{2}r^2}
	{4}+m^2.
\end{eqnarray}
The Laplacian operator in cylindrical coordinates reads $\bs{\nabla}^{2}=\partial_{r}^{2}+\frac{1}{r}\partial_r+\frac{1}{r^2}\partial_{\theta}^2+\partial_{z}^{2}$. The solution of \eqref{A6}
\begin{eqnarray}\label{A8}
\varphi_{n\ell}(x,k)&=&C_{n\ell}e^{-iEt+i\ell\theta+ik_{z}z}\nonumber\\
&&\times e^{-eB_{0}r^2/4}r^{\abs{\ell}}L_{n}^{\abs{\ell}}\left(\frac{eB_0r^2}{2}\right),
\end{eqnarray}
is given in terms of the associated Laguerre polynomials $L_{n}^{m}(u)$, with the normalization factor
\begin{eqnarray}\label{A9}
C_{n\ell}=\left(\frac{(eB_{0})^{|\ell|+1}}{2^{|\ell|}}\frac{n!}{(n+\ell)!}\right)^{1/2}.
\end{eqnarray}
In \eqref{A8}, $n\in\mathbb{N}_{0}$ labels the Landau levels and $\ell\in\mathbb{Z}$ is the quantum number associated with the operator $L_{z}$.
\par
The free propagator of the magnetized and rotating KG fields, $\mathcal{G}(x,x')$, can be derived by using the generalized Fock-Schwinger formalism \cite{siri2024-1, abedlou2025}. In coordinate space, it is expressed in terms of the eigenfunctions $\varphi_{\Lambda}$ of the differential operator $\mathfrak{D}(x,\partial_{x})$ from \eqref{A7},
\begin{eqnarray}\label{A10}
\hspace{-0.5cm}\mathcal{G}(x,x')=-i\int_{-\infty}^{.0} d\tau \sum_{\Lambda}e^{-i\Lambda\tau}\varphi_{\Lambda}(x)\varphi^{*}_{\Lambda}(x').
\end{eqnarray}
Here, $\tau$ denotes the proper-time and $\Lambda$ in $e^{-i\Lambda\tau}$ represents the eigenvalue of the differential operator $\mathfrak{D}(x,\partial_x)$. Plugging \eqref{A8} into \eqref{A10}, using $$\Lambda=-\tilde{E}^2+k_z^{2}+(2n+1)eB_0+m^{2},$$ performing the integration over $\tau$, and a change of variable $E\to E-\ell\Omega$, we first arrive at
\begin{widetext}
\begin{eqnarray}\label{A11}
\mathcal{G}(x,x')&=&\sum_{n=0}^{\infty}\sum_{\ell=-\infty}^{+\infty}\int\frac{d{E}d{k_z}}{(2\pi)^2}\abs{C_{n\ell}}^{2}\frac{e^{-iE(t-t')+i\ell\Omega(t-t')+i\ell(\theta-\theta')+ik_z(z-z')}}{-E^2+k_z^{2}+m^2+(2n+1)eB_0} \nonumber\\
&\times& e^{-eB_0r^2/4}r^{\abs{\ell}}L_{n}^{\abs{\ell}}\left(\frac{eB_0r^2}{2}\right)e^{-eB_0r'^2/4}r'^{\abs{\ell}}L_{n}^{\abs{\ell}}\left(\frac{eB_0r'^2}{2}\right).
\end{eqnarray}
\end{widetext}
Then, plugging \eqref{A8} and \eqref{A11} into
\begin{eqnarray}\label{A12}
	\tilde{\mathcal{G}}_{n\ell,n'\ell'}(k,k')=\int_{x,x'}\mathcal{G}(x,x')\varphi_{n\ell}^{*}(x,k)\varphi_{n'\ell'}(x',k'),\nonumber\\
\end{eqnarray}
where
\begin{eqnarray}\label{A13}
	\int_{x,x'}\equiv\int d^{4}{x}d^{4}{x'},\quad\mbox{with}\quad d^{4}{x}=d{t}\ rd{r}\ d{\theta}\ d{z}, \nonumber\\
\end{eqnarray}
and performing the integration over $t, r, \theta,$ and $z$ by utilizing, in particular, the orthonormality relation
of the Laguerre polynomials
\begin{eqnarray*}
\int_{0}^{\infty}dx x^{\ell}e^{-x} L_{n}^{\ell}(x)L_{m}^{\ell}(x)=\frac{(n+\ell)!}{n!}\delta_{nm},
\end{eqnarray*}
we arrive at the free propagator in the momentum space,
\begin{eqnarray}\label{A14}
	\tilde{\mathcal{G}}_{n\ell,n'\ell'}(k,k')=(2\pi)^4\widehat{\Delta}_{n\ell,n'\ell'}(k,k') \mathcal{G}_{n\ell}(k).
\end{eqnarray}
Here,
\begin{eqnarray}\label{A15}
	\widehat{\Delta}_{n\ell,n'\ell'}(k,k')\equiv \delta(k_0-k'_0)\delta(k_z-k'_z)\delta_{\ell\ell'}\delta_{nn'},
\end{eqnarray}
and
\begin{eqnarray}\label{A16}
\mathcal{G}_{n\ell}(k)\equiv \frac{1}{\omega_{B}^{2}-\left(k_0+\ell\Omega\right)^2}.
\end{eqnarray}
Moreover, the energy dispersion relation is given by $\omega_{B}^{2}\equiv k_z^2+eB_0\left(2n+1\right)+m^2$. In what follows, we determine the thermodynamic potential $\Phi$.  To do this, we introduce temperature by replacing $k_{0}$ with $i\omega_{k} + \mu$, where $\omega_{k}$ represents the Matsubara frequency defined as $\omega_{k} \equiv 2\pi k T$ with $k \in \mathbb{Z}$, $T$ being the temperature, and $\mu$ the chemical potential. The boson propagator at finite temperature and density is thus given by
\begin{eqnarray}\label{A17}
\mathcal{G}_{n\ell}(\omega_{k},k_{z})=\frac{1}{\omega_{B}^{2}+\left(\omega_{k}+i\mu_{\text{eff}}\right)^{2}},
\end{eqnarray}
where the effective chemical potential is defined by $\mu_{\text{eff}}\equiv \mu+\mu_{\ell}$ with $\mu_{\ell}\equiv \ell\Omega$.
\subsection{Thermodynamical potential}\label{sec2B}
The thermodynamic potential of this model is determined by plugging the bosonic propagator \eqref{A17} into
\begin{eqnarray}\label{A18}
\Phi=\frac{T}{2S}\sum_{k=-\infty}^{\infty}\sum_{n=0}^{\infty}\sum_{\ell=-n}^{N-n}\int \frac{dk_{z}}{2\pi}\ln\left[\beta^2 \mathcal{G}_{n\ell}^{-1}(\omega_{k},k_{z})\right].\nonumber\\
\end{eqnarray}
Here, we used
\begin{eqnarray}\label{A19}
\int\frac{dk_{0}}{2\pi}\to iT\sum_{k},\qquad \int\frac{dk_{x}dk_{y}}{(2\pi)^{2}}\to \frac{1}{S}\sum_{n=0}^{\infty}\sum_{\ell=-n}^{N-n}, \nonumber\\
\end{eqnarray}
where $S\equiv \pi \bar{R}^{2}$ with $\bar{R}$ denoting the radius of the hypothetical cylinder that encloses the system (see \cite{fukushima2015} for further details). Here, in contrast to the case with $B=0$, the summation over $\ell$ has an upper bound $N-n$, where $N$ is the Landau degeneracy factor defined by
\begin{eqnarray}\label{A20}
N\equiv \lfloor\frac{eB_{0}S}{2\pi} \rfloor.
\end{eqnarray}
The summation over Matsubara frequencies are performed by making use of
\begin{eqnarray}\label{A21}
\sum_{n=-\infty}^{+\infty}\ln[(2\pi n)^{2}+u^{2}]=u+2\ln\left(1-e^{-u}\right).
\end{eqnarray}
Focusing only on the $T$-dependent part of the potential, $\Phi_{T}$, we obtain
\begin{eqnarray}\label{A22}
\Phi_{T}=\Phi_{T}^{+}+\Phi_{T}^{-},
\end{eqnarray}
with
\begin{eqnarray}\label{A23}
\Phi_{T}^{\pm}\equiv\frac{T}{2S}\sum_{n=0}^{\infty}\sum_{\ell=-n}^{N-n}\int \frac{dk_{z}}{2\pi}\ln\left(1-e^{-\beta\left(\omega_{B}\mp\mu_{\text{eff}}\right)}\right).\nonumber\\
\end{eqnarray}
Here, the superscripts  $+$ and $-$ correspond to particle and antiparticle contributions to the thermodynamic potential, respectively. To study the Bose-Einstein condensation of particles, we neglect the antiparticle part of $\Phi_{T}$. Moreover, we consider only the nonrelativistic limit, where $\omega_{B}$ is approximated by
\begin{eqnarray}\label{A24}
\omega_{B}\approx \omega+m_{B}, \quad\text{with}\quad\omega\equiv \frac{k_{z}^{2}+2neB_{0}}{2m_{B}},
\end{eqnarray}
and $m_{B}\equiv \left(m^{2}+eB_{0}\right)^{1/2}$. Plugging \eqref{A24} into $\Phi_{T}^{+}$, we arrive first at
\begin{eqnarray}\label{A25}
\Phi_{T}^{+}=
\frac{T}{2S}\sum_{n=0}^{\infty}\sum_{\ell=-n}^{N-n}\int\frac{dk_{z}}{2\pi}\ln\left(1-e^{\beta\mu_{B}}e^{-\beta\left(\omega-\mu_{\ell}\right)}\right),\nonumber\\
\end{eqnarray}
with $\mu_{B}\equiv \mu-m_{B}$. 
At this stage, we need to define an appropriate cutoff to ensure that the argument of the logarithm function appearing in \eqref{A25} remains positive. To do this, we modify
$\mu_{\ell}$ by replacing it with $\mu_{\ell}-\mu_{\Omega}$, where $\mu_{\Omega}\equiv (N-n)\Omega$. Assuming that $\mu_{B}<0$ and noting that $\omega$ is always positive for $n\in\mathbb{N}_{0}$, we can conclude that the remaining exponential satisfies
$
e^{\beta (\mu_{\ell}-\mu_{\Omega})}\leq 1
$ for all values of $\ell\in\{-n,N-n\}$. Introducing the ``magnetorotational fugacity'',
\begin{eqnarray}\label{A26}
z_{B,\Omega}^{(n)}\equiv e^{\beta\left(\mu_{B}-\mu_{\Omega}\right)},
\end{eqnarray}
the thermodynamic potential of particles thus reads
\begin{eqnarray}\label{A27}
\Phi_{T}^{+}=
\frac{T}{2S}\sum_{n=0}^{\infty}\sum_{\ell=-n}^{N-n}\int\frac{dk_{z}}{2\pi}\ln\left(1-z_{B,\Omega}^{(n)}e^{-\beta\left(\omega-\mu_{\ell}\right)}\right). \nonumber\\
\end{eqnarray}
To evaluate \eqref{A27}, we use
\begin{eqnarray}\label{A28}
\ln(1-x)=-\sum_{j=1}^{\infty}\frac{x^{j}}{j},
\end{eqnarray}
to arrive first at
\begin{eqnarray}\label{A29}
\Phi_{T}^{+}&=&-\frac{T}{2S}\sum_{j=1}^{\infty}\frac{\left(z_{B,\Omega}^{(n)}\right)^{j}}{j}\sum_{n=0}^{\infty}e^{-\frac{\beta jeB_{0}n}{m_{B}}}\sum_{\ell=-n}^{N-n}e^{\beta j\mu_{\ell}}\nonumber\\
&&\times \int\frac{dk_{z}}{2\pi}e^{-\frac{\beta jk_{z}^{2}}{2m_{B}}}.
\end{eqnarray}
The integration over $k_{z}$ yields
\begin{eqnarray}\label{A30}
 \int_{-\infty}^{+\infty}\frac{dk_{z}}{2\pi}e^{-\frac{\beta jk_{z}^{2}}{2m_{B}}}=\frac{1}{j^{1/2}\lambda_{B}},
\end{eqnarray}
with the `'thermomagnetic length'' defined by
\begin{eqnarray}\label{A31}
\lambda_{B}\equiv\left(\frac{2\pi}{m_{B}T}\right)^{1/2}.
\end{eqnarray}
We then perform the summation over $\ell$ in \eqref{A29} by replacing it with an appropriate integral and assuming $\beta\Omega\ll 1$,
\begin{eqnarray}\label{A32}
\sum_{\ell=-n}^{N-n}e^{\beta j\mu_{\ell}}\simeq \frac{1}{\Omega}\int_{-n\Omega}^{(N-n)\Omega}d\mu_{\ell}~e^{\beta j\mu_{\ell}}\approx N.
\end{eqnarray}
As concerns the summation over Landau levels $n$, we separate the contribution from the LLL from that from the HLL. By utilizing
\begin{eqnarray}\label{A33}
\sum_{n=1}^{\infty}e^{-\frac{\beta jeB_{0}n}{m_{B}}}=\frac{1}{\exp\left(\frac{\beta jeB_{0}}{m_{B}}\right)-1},
\end{eqnarray}
and plugging \eqref{A30} and \eqref{A32} into \eqref{A29}, we arrive at
\begin{eqnarray}\label{A34}
\Phi_{T}^{+}=\Phi_{T}^{\text{LLL}}+\Phi_{T}^{\text{HLL}},
\end{eqnarray}
with LLL ($n=0$) part of  $\Phi_{T}^{+}$
\begin{eqnarray}\label{A35}
\Phi_{T}^{\text{LLL}}=-\frac{eB_{0}T}{2\pi\lambda_{B}}\text{Li}_{3/2}\left(z_{B,\Omega}^{(0)}\right),
\end{eqnarray}
and its HLL ($n\geq 1$) contributions
\begin{eqnarray}\label{A36}
\Phi_{T}^{\text{HLL}}=-\frac{2B_{0}T}{2\pi\lambda_{B}}\sum_{j=1}^{\infty}\frac{\left(z_{B,\Omega}^{(n)}\right)^{j}}{j^{3/2}}\frac{1}{\exp\left(\frac{\beta jeB_{0}}{m_{B}}\right)-1}. \nonumber\\
\end{eqnarray}
In \eqref{A35}, the polylogarithm function is defined by
\begin{eqnarray}\label{A37}
\text{Li}_{\nu}(x)\equiv \sum_{j=1}^{\infty}\frac{x^{j}}{j^{\nu}}.
\end{eqnarray}
In Sec. \eqref{sec3}, we use $\Phi_{T}^{\text{LLL}}$ to explore a potential BEC scenario in the presence of rotation and magnetic field. The thermodynamic properties of rotating bosons subjected to a constant magnetic field are examined in Sec. \ref{sec4}.
\section{Weak scenario of BEC in the presence of $\bs{\Omega}$ and $\bs{B}$}\label{sec3}
\setcounter{equation}{0}
In this section, we present the weak scenario of BEC by using the method introduced in \cite{rojas2025}. We focus solely on the contribution of LLL to the thermodynamic potential $\Phi_{T}^{+}$. We begin by calculating the number density of rotating and magnetized bosons through the thermodynamic relation
\begin{eqnarray}\label{B1}
n_{\text{th}}=-\left(\frac{\partial\Phi_{T}^{\text{LLL}}}{\partial \mu}\right)_{T,B,\Omega}.
\end{eqnarray}
Plugging $\Phi_{T}^{\text{LLL}}$ from \eqref{A35}, and utilizing
\begin{eqnarray}\label{B2}
\frac{\partial \mbox{Li}_{\nu}(x)}{\partial x}=\frac{1}{x}\mbox{Li}_{\nu-1}(x),
\end{eqnarray}
we arrive at\footnote{For simplicity, we omit the superscript $(0)$ on $z_{B,\Omega}^{(0)}$ that appears in \eqref{A35}.}
\begin{eqnarray}\label{B3}
n_{\text{th}}=\frac{eB_{0}}{2\pi \lambda_{B}}\mbox{Li}_{1/2}\left(z_{B,\Omega}\right).
\end{eqnarray}
Let us remember that the number density for $\Omega=0$ and $B=0$ is given by \cite{kapusta-book}
\begin{eqnarray}\label{B4}
n_{\text{th}}\big|_{\Omega=B=0}=\frac{1}{\lambda_{T}^{3}}\mbox{Li}_{3/2}(z),
\end{eqnarray}
with $\lambda_{T}=\left(\frac{2\pi}{mT}\right)^{1/2}$, while for $\Omega\neq 0$ and $B=0$, it reads \cite{siri2024-2}
\begin{eqnarray}\label{B5}
n_{\text{th}}\big|_{\Omega\neq 0,B=0}=\frac{1}{\lambda_{T,\Omega}^{3}}\mbox{Li}_{5/2}(z), 
\end{eqnarray}
with $\lambda_{T,\Omega}=\lambda_{T}(\beta\Omega)^{1/3}$. In \eqref{B4} and \eqref{B5}, the fugacity is defined by $z\equiv e^{\beta(\mu-m)}$.
The dimensional reduction resulting from the presence of a constant $B$ field is clearly evidenced by the order of the polylogarithm functions appearing in \eqref{B3} and \eqref{B4}.
\par
In typical BEC scenarios, condensation occurs at temperatures below a specific critical temperature  $T_{c}$. At these temperatures, the generic fugacity equals $1$, indicating a strong BEC scenario \cite{rojas2025}. As the temperature decreases, an increasing number of bosons occupy the ground state. When the temperature approaches zero, all bosons in the gas occupy the ground state.
\par
As it is described in \cite{rojas2025}, however, the presence of external magnetic fields leads to a weak scenario of BEC. This scenario is characterized by three distinct properties:
\begin{enumerate}
\item A critical temperature cannot be defined. Instead, the system exhibits a diffuse phase transition over specific temperature intervals. In this situation, bosons tend to concentrate around the ground state, contrasting the strong BEC scenario.
\item The fugacity remains less than $1$ at all temperatures.
\item There is no peak in the specific heat of the Bose gas.
\end{enumerate}
In the present paper, we discuss a scenario where, aside from the influence of $B$, the system is in a state of rigid rotation. We observe that a similar weak scenario arises in this context. To demonstrate this, let us first set $z_{B,\Omega} = 1$ in \eqref{B3}. Utilizing
$\mbox{Li}_{\nu}(1) = \zeta(\nu)$, where $\zeta(\nu)$ represents the Riemann zeta function, we obtain
\begin{eqnarray}\label{B6}
n_{\text{th}}(z_{B,\Omega}=1)=\frac{eB_{0}}{2\pi\lambda_{B}}\zeta\left(1/2\right).
\end{eqnarray}
This result is unacceptable because $\zeta(1/2)\simeq -1.46$ leads to a negative number density. To study the weak BEC scenario in this case, we separate the number density $n_{\text{th}}$ into two parts: the ground state part, $n_{\text{gr}}$, and the excited part $n_{\text{ex}}$,
\begin{eqnarray}\label{B7}
n_{\text{th}}=n_{\text{gr}}+n_{\text{ex}}.
\end{eqnarray}
To determine $n_{\text{gr}}$, we return to \eqref{A29}, set $n=0$, and specifically introduce a finite cutoff $p_{0}$ for the integration over $k_{z}$. Using
\begin{eqnarray}\label{B8}
 \int_{-p_{0}}^{+p_{0}}\frac{dk_{z}}{2\pi}e^{-\frac{\beta jk_{z}^{2}}{2m_{B}}}=\frac{1}{j^{1/2}\lambda_{B}}\text{Erf}\left(\left(\frac{\beta j p_{0}^{2}}{2m_{B}}\right)^{1/2}\right). \nonumber\\
\end{eqnarray}
After performing the summation over $\ell\in\{0,N\}$ using \eqref{A32}, the ground state part $\Phi_{\text{LLL}}^{+}$ is given by
\begin{eqnarray}\label{B9}
\Phi_{T}^{+/\text{gr}}= -\frac{eB_{0}T}{2\pi \lambda_{B}}\sum_{j=1}^{\infty}\frac{\left(z_{B,\Omega}\right)^{j}}{j^{3/2}}\text{Erf}\left(\left(\frac{\beta j p_{0}^{2}}{2m_{B}}\right)^{1/2}\right). \nonumber\\
\end{eqnarray}
Plugging \eqref{B9} into \eqref{B1}, the ground state part of the number density reads
\begin{eqnarray}\label{B10}
\hspace{-0.5cm}
n_{\text{gr}}=\frac{eB_{0}}{2\pi \lambda_{B}}\sum_{j=1}^{\infty}\frac{\left(z_{B,\Omega}\right)^{j}}{j^{1/2}}\text{Erf}\left(\left(\frac{\beta j p_{0}^{2}}{2m_{B}}\right)^{1/2}\right).
\end{eqnarray}
As $T\to 0$, independently of $p_{0}$, both $n_{\text{gr}}\to n_{\text{th}}$ and $n_{\text{ex}}\to 0$. Hence, as stated in \cite{rojas2025}, there exists a temperature range in which all the particles are confined within any arbitrary neighborhood of the ground state. This phenomenon accounts for the first characteristic of the weak BEC scenario.
\par
As concerns the second characteristic of this scenario, we need to determine the $T$ dependence of $z_{B,\Omega}$. This dependence arises from the chemical potential $\mu$. To establish the $T$ dependence of $\mu$, we require that $n_{\text{th}}$ remains constant,
\begin{eqnarray}\label{B11}
\hspace{-0.5cm}n_{\text{th}}\left(T,eB_{0},\mu\left(T,eB_{0},\Omega,\rho_{0}\right)\right)=\rho_{0}=\text{const.}.
\end{eqnarray}
Plugging \eqref{B3} into \eqref{B11} and using the asymptotic expansion of $\text{Li}(e^{-\alpha})$ for $\alpha\ll 1$
\begin{eqnarray}\label{B12}
	\mbox{Li}_{1/2}\left(e^{-\alpha}\right)\simeq\sqrt{\frac{\pi}{\alpha}}+\zeta(1/2),
\end{eqnarray}
with $\alpha\equiv\beta\left(m_B+N\Omega-\mu\right)$, we arrive at
\begin{eqnarray}\label{B13}
\rho_{0}\simeq \frac{eB_{0}}{2\pi\lambda_{B}}\left[\sqrt{\frac{\pi}{\alpha}}+\zeta(1/2)\right],
\end{eqnarray}
that leads eventually to
\begin{eqnarray}\label{B14}
\hspace{-0.5cm}
\mu\simeq N\Omega+m_B-\pi T \left[\frac{2\pi\rho_{0}\lambda_{B}}{eB_0}-\zeta(1/2)\right]^{-2}.
\end{eqnarray}
We observe that the assumption $\alpha \ll 1$ is either equivalent to high temperatures or to a large value of $\mu_{B}-N\Omega$. These assumptions are necessary and lead to $z_{B,\Omega} < 1$, which is required in the weak scenario of BEC. Plugging \eqref{B14} into $z_{B,\Omega}=e^{-\alpha}$, it is approximately given by
\begin{eqnarray}\label{B15}
z_{B,\Omega}\simeq \exp\left(-\pi\left[\frac{2\pi\rho_{0}\lambda_{B}}{eB_0}-\zeta(1/2)\right]^{-2}\right).
\end{eqnarray}
In this approximation, $z_{B,\Omega}<1$ for all values of $T$, as expected. The fugacity is, however, independent of $\Omega$ and its $T$ dependence arises solely from $\lambda_{B}$. To restore the $\Omega$ dependence, we must remember that the thermodynamic parameters $T,\mu$, and $\Omega$ are defined in the corotating frame. Here, we need to consider the Tolman-Ehrenfest factor $\Gamma(v)\equiv 1/\sqrt{1-v^{2}}$, with the linear velocity $v\equiv R\Omega$ to distinguish these parameters from those in the initial frame \cite{tolman1930, ehrenfest1930, chernodub2012}. In Sec. \ref{sec5}, we replace $T$ in our numerical computations with $\Gamma(v) T$ to account for the relativistic correction to temperature and its dependence on the distance $R$ from the rotation axis for a fixed $\Omega$. In what follows, before presenting the numerical results, particularly the $T$ dependence of $n_{\text{gr}}$, we will focus on the thermodynamic quantities derived from $\Phi_{T}^{\text{LLL}}$ and provide analytical expressions of these quantities in Sec. \ref{sec4}. In Sec. \ref{sec5}, we will then explore their dependence on $T,B$ and $\Omega$.
\begin{figure}[h]
\includegraphics[width=8cm, height=6cm]{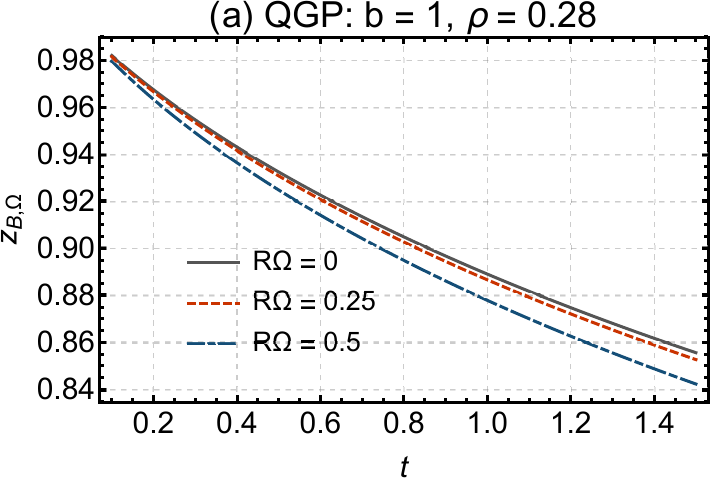}\vspace{0.5cm}
\includegraphics[width=8cm, height=6cm]{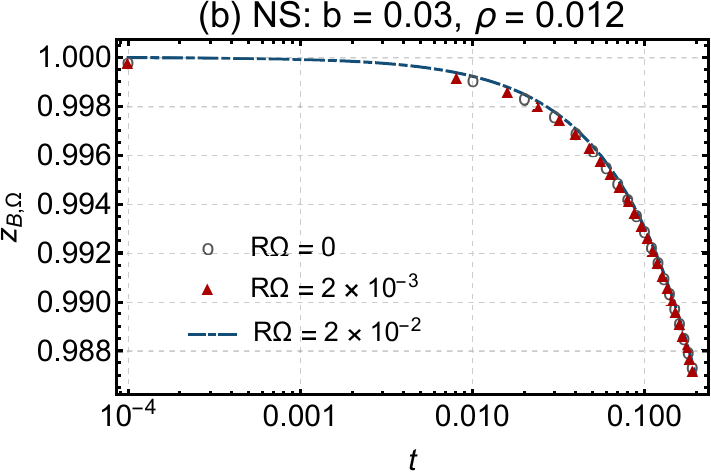}
\caption{Panel a: The temperature ($t$) dependence of the magnetorotational fugacity $z_{B,\Omega}$ is plotted for the QGP case and various linear velocities $R\Omega$. In the relevant regime $0.2<t<1.5$ for the QGP case, $z_{B,\Omega}$ is less than $1$ and decreases as $v$ increases.
 Panel b: The $t$ dependence of $z_{B,\Omega}$ is plotted for the NS case and different linear velocities $R\Omega$. Within the relevant regime $10^{-4}<t<0.1$ for the NS case, varying $R\Omega$ does not significantly alter the results. In both cases, the dependence on $v$ arises from the Tolman-Ehrenfest factor.}\label{fig1}
\end{figure}
\begin{figure}[h]
\includegraphics[width=8cm, height=6cm]{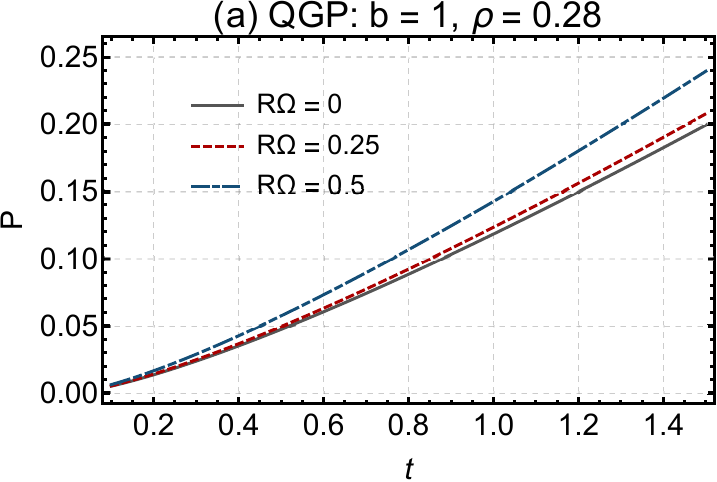}\vspace{0.5cm}
\includegraphics[width=8cm, height=6cm]{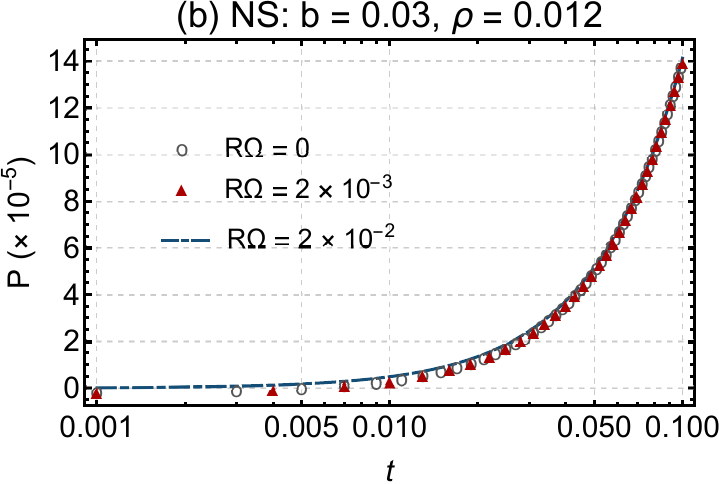}
\caption{The temperature ($t$) dependence of the pressure $P=-\Phi_{T}^{\text{LLL}}$, with the LLL contribution to the thermodynamic potential, $\Phi_{T}^{\text{LLL}}$ is plotted for the QGP (panel a) and the NS (panel b) cases and various linear velocities $R\Omega$. Here, we used the data of Fig. \ref{fig1}. Whereas the QGP pressure increases with increasing $t$ and $\Omega$, the values of $R\Omega$ do not significantly change the NS pressure results. }\label{fig2}
\end{figure}
\section{Thermodynamic quantities: Analytical results}\label{sec4}
\setcounter{equation}{0}
In this section, we explore the $T,B,$ and $\Omega$ dependence of several thermodynamic quantities arising from the thermodynamic potential $\Phi_{T}^{\text{LLL}}$ from \eqref{A35}. We present analytical expressions for the angular-momentum density $j$, magnetization $M$, entropy density $s$, and energy density $\epsilon$ in the corotating frame defined by
\begin{eqnarray}\label{C1}
j&\equiv& -\left(\frac{\partial \Phi_{T}^{\text{LLL}}}{\partial \Omega}\right)_{\mu,T,B,v},\quad\epsilon\equiv -T^{2}\left(\frac{\partial}{\partial T}\left(\frac{\Phi_{T}^{\text{LLL}}}{T}\right)\right)_{z_{B}},\nonumber\\
s&\equiv& -\left(\frac{\partial \Phi_{T}^{\text{LLL}}}{\partial T}\right)_{\mu,\Omega,B},\qquad
\hspace{-0.5cm}M\equiv -e\left(\frac{\partial \Phi_{T}^{\text{LLL}}}{\partial \left(eB\right)}\right)_{\mu,T,\Omega}.\nonumber\\
\end{eqnarray}
We also determine the specific heat $C_{V}$, moment of inertia $I$, free susceptibility $\chi_{f}$, magnetic susceptibility $\chi_{m}$,
\begin{eqnarray}\label{C2}
C_{V}&\equiv&\left(\frac{\partial\epsilon}{\partial T}\right)_{n_{\text{th}},B,\Omega},\qquad I\equiv\left(\frac{\partial j}{\partial\Omega}\right)_{\mu,T,B,v}, \nonumber\\
\chi_{f}&\equiv& T\left(\frac{\partial n_{\text{th}}}{\partial\mu}\right)_{T,B,\Omega},\qquad \hspace{-0.2cm}\chi_{m}\equiv e\left(\frac{\partial M}{\partial \left(eB\right)}\right)_{\mu,T,\Omega}, \nonumber\\
\end{eqnarray}
and isothermal compressibility $\kappa_{T}$,
\begin{eqnarray}\label{C3}
\kappa_{T}\equiv \frac{1}{n_{\text{th}}}\left(\frac{\partial n_{\text{th}}}{\partial P}\right)_{T,B,\Omega}=\frac{\chi_{f}}{n^{2}_{}T}.
\end{eqnarray}
We begin with the angular-momentum density $j$. Plugging \eqref{A35} into $j$ from \eqref{C1} and using \eqref{appB1}, we arrive at
\begin{eqnarray}\label{C4}
j= -\frac{eB_0N}{2\pi\lambda_{B}}\mbox{Li}_{1/2}(z_{B,\Omega}).
\end{eqnarray}
Comparing this result with \eqref{B3} for $n_{\text{th}}$, we obtain $j=-Nn_{\text{th}}$.
\par
Using \eqref{C1}, the energy density in the corotating frame reads,
\begin{eqnarray}\label{C5}
\epsilon=\frac{P}{2}-j\Omega,
\end{eqnarray}
with the pressure $P\equiv -\Phi_{T}^{\text{LLL}}$ and $\Phi_{T}^{\text{LLL}}$ from \eqref{A35} and $j$ from \eqref{C4}. Plugging \eqref{C5} into the equation for the energy density in the inertial (laboratory) frame,
 $\epsilon^{\text{lab}}=\epsilon+j\Omega$, we find that  $\epsilon^{\text{lab}}=P/2$. Using this equation of state in the laboratory frame, the sound velocity in this frame reads $c_{s}^{\text{lab}}=\partial P/\partial \epsilon^{\text{lab}}=\sqrt{2}\sim 1.4$. It is important to note that in the absence of $\Omega$ and $B$, the speed of sound in the NR limit is given by $c_{s}=\sqrt{2/3}\sim 0.82$. This value arises from the equation of state $\epsilon^{\text{lab}}=3P/2$. The difference between $\epsilon^{\text{lab}}=P/2$ for $B\neq 0$ and $\Omega\neq 0$ and $\epsilon^{\text{lab}}=3P/2$ for $B=0$ and $\Omega=0$ illustrates the dimensional reduction caused by the presence of the $B$ field, as discussed in Sec. \ref{sec3}. Additionally, as shown in \cite{siri2024-2}, for a nonrelativistic and rigidly rotating Bose gas without the $B$ field, the equation of state in the corotating frame is $\epsilon=5P/2$, whereas in the laboratory frame it remains $\epsilon^{\text{lab}}=3P/2$.
\par
By substituting $\Phi_{T}^{\text{LLL}}$ into the definition of the entropy density $s$ in \eqref{C1}, we obtain
\begin{eqnarray}\label{C6}
s=\beta\left(\epsilon+P-\mu_{B}n_{\text{th}}\right)=\mathfrak{s}+\beta m_{B}n_{\text{th}},
\end{eqnarray}
where the definition of the thermodynamic entropy $\mathfrak{s}\equiv \beta\left(\epsilon+P-\mu n_{\text{th}}\right)$ and $\mu_{B}=\mu-m_{B}$ are used. The distinction between $s$ and $\mathfrak{s}$ arises from
the way the magnetorotational fugacity $z_{B,\Omega}$ is defined in terms of $m_{B}$. This difference is also noted in \cite{siri2024-2}.
\par
Additionally, the magnetization $M/e$ derived from \eqref{C3} is given by
\begin{eqnarray}\label{C7}
M&=&\frac{eT}{2\pi\lambda_{B}}\left[\mbox{Li}_{3/2}(z_{B,\Omega})+\frac{eB_0}{4m_B^2}\mbox{Li}_{3/2}(z_{B,\Omega})\right.\nonumber\\
&&\left.-\frac{eB_0}{2m_BT}\mbox{Li}_{1/2}(z_{B,\Omega})\right].
\end{eqnarray}
As concerns $C_{V}$ from \eqref{C2}, we assume $\partial n_{\text{th}}/\partial T=0$ and arrive first at
\begin{eqnarray}\label{C8}
\left(\frac{\partial z_{B,\Omega}}{\partial T}\right)_{n_{\text{th}},B,\Omega}=-\frac{z_{B,\Omega}}{2T}\frac{\mbox{Li}_{1/2}(z_{B,\Omega})}{\mbox{Li}_{-1/2}(z_{B,\Omega})}.
\end{eqnarray}
Then, plugging $\epsilon$ from \eqref{C5} into the definition of $C_{V}$ from \eqref{C2} and using \eqref{C8}, we obtain
\begin{eqnarray}\label{C9}
C_{V}=\frac{eB_0}{2\pi\lambda_{B}}\left[\frac{3}{4}\mbox{Li}_{3/2}(z_{B,\Omega})-\frac{1}{4}\frac{\mbox{Li}_{1/2}^{2}(z_{B,\Omega})}{\mbox{Li}_{-1/2}(z_{B,\Omega})}\right].\nonumber\\
\end{eqnarray}
From this, the ratio $C_{V}/n_{\text{th}}$ is expressed as
\begin{eqnarray}\label{C10}
	 \frac{C_V}{n_{\text{th}}}=\frac{3}{4}\frac{\mbox{Li}_{3/2}(z_{B,\Omega})}{\mbox{Li}_{1/2}(z_{B,\Omega})}-\frac{1}{4}\frac{\mbox{Li}_{1/2}(z_{B,\Omega})}{\mbox{Li}_{-1/2}(z_{B,\Omega})}.
\end{eqnarray}
In the high-temperature limit, $\mbox{Li}_{\nu}(x)\simeq x$. Thus, we have
\begin{eqnarray}\label{C11}
\lim\limits_{T\to \infty} C_V\simeq \frac{n_{\text{th}}}{2}.
\end{eqnarray}
The moment of inertia $I$ is given by plugging $j$ from \eqref{C4} into its definition from \eqref{C2},
\begin{eqnarray}\label{C12}
I=\frac{eB_{0}N^{2}}{2\pi\lambda_{B}T}\mbox{Li}_{-1/2}\left(z_{B,\Omega}\right).
\end{eqnarray}
Similarly, the free susceptibility of a rigidly rotating magnetized Bose gas is given by
\begin{eqnarray}\label{C13}
\chi_{f}=\frac{eB_{0}}{2\pi\lambda_{B}}\mbox{Li}_{-1/2}\left(z_{B,\Omega}\right).
\end{eqnarray}
Here, we utilize \eqref{B2} and the definition of $\chi_{f}$ from \eqref{C2}. By comparing $\chi_{f}$ with $I$ from \eqref{C12}, we obtain
\begin{eqnarray}\label{C14}
\chi_{f}=\frac{T}{N^{2}}I.
\end{eqnarray}
Plugging $M$ from \eqref{C7} into the definition of the magnetic susceptibility $\chi_{m}$ from \eqref{C2}, the magnetic susceptibility reads
\begin{eqnarray}\label{C15}
\chi_{m}&=&\frac{e^{2}T}{2\pi\lambda_{B}}\bigg[\frac{1}{2m_B^2}\mbox{Li}_{3/2}(z_{B,\Omega})-\frac{3eB_{0}}{16m_{B}^{4}}\mbox{Li}_{3/2}(z_{B,\Omega})\nonumber\\
&&-\frac{1}{m_{B}T}\mbox{Li}_{1/2}(z_{B,\Omega})+\frac{eB_{0}}{4m_{B}^{2}T^{2}}\mbox{Li}_{-1/2}(z_{B,\Omega})\bigg].\nonumber\\
\end{eqnarray}
For the  numerical results, we use $e^{2}=4\pi\alpha$, where $\alpha=1/137$ is the fine-structure constant.
Finally, substituting $\chi_{f}$ from \eqref{C14} and $n_{\text{th}}$ from \eqref{B3} into \eqref{C3}, we arrive first at
\begin{eqnarray}\label{C16}
\kappa_{T}=\frac{I}{N^{2}n_{\text{th}}^{2}}.
\end{eqnarray}
Alternatively, we can express this as
\begin{eqnarray}\label{C17}
\kappa_{T}=\frac{2\pi\lambda_{B}}{eB_{0}T}\frac{\mbox{Li}_{-1/2}\left(z_{B,\Omega}\right)}{\mbox{Li}_{1/2}^{2}\left(z_{B,\Omega}\right)}.
\end{eqnarray}
It would be intriguing to compare this result with the corresponding expressions for two cases: nonrotating and rotating Bose gases in the absence of a magnetic field (i.e., $\Omega=0, B=0$ and $\Omega\neq 0, B=0$).\footnote{Following the same reasoning, all other quantities presented in this section can be compared with the corresponding quantities in these two cases.}
\par
As it turns out, for the case where $\Omega=0, B=0$, the free susceptibility of a nonrelativistic free Bose gas is given by
\begin{eqnarray}\label{C18}
\chi_{f}\big|_{\Omega=B=0}=\frac{1}{\lambda_{T}^{3}}\mbox{Li}_{1/2}(z),
\end{eqnarray}
where $\lambda_{T}$ is defined in \eqref{B4}. Plugging $\chi_{f}$ from \eqref{C18} and $n_{\text{th}}$ from \eqref{B4} into \eqref{C3}, we find the isothermal compressibility as
\begin{eqnarray}\label{C19}
\kappa_{T}\big|_{\Omega=B=0}=\frac{\lambda_{T}^{3}}{T}\frac{\mbox{Li}_{1/2}(z)}{\mbox{Li}_{3/2}^{2}(z)}.
\end{eqnarray}
For the case where $\Omega\neq 0, B=0$, we arrive at the free susceptibility of a nonrelativistic rotating bosonic plasma
\begin{eqnarray}\label{C20}
\chi_{f}\big|_{\Omega\neq 0, B=0}=\frac{1}{\lambda_{T,\Omega}^{3}}\mbox{Li}_{3/2}(z),
\end{eqnarray}
where $\lambda_{T,\Omega}$ is defined in \eqref{B5}. By substituting $\chi_{f}$ from \eqref{C20} and $n_{\text{th}}$ from \eqref{B5} into \eqref{C3}, the isothermal compressibility in this case is expressed as
\begin{eqnarray}\label{C21}
\kappa_{T}\big|_{\Omega\neq 0,B=0}=\frac{\lambda_{T,\Omega}^{3}}{T}\frac{\mbox{Li}_{3/2}(z)}{\mbox{Li}_{5/2}^{2}(z)}.
\end{eqnarray}
It is important to note that the results \eqref{C14} and \eqref{C16}, which express the relation between $\chi_{f}$ and $\kappa_{T}$ on the moment of inertia $I$, are novel. At this stage, two remarks regarding these quantities are in order. According to \eqref{C14}, the free susceptibility of the magnetized and rotating Bose gas depends on three main physical quantities, $T, N$, and $I$. The linear dependence on $T$ indicates the statistical origin of this response. The free susceptibility increases with rising temperature. Additionally, as the temperature increases, thermal fluctuations ($n_{\text{th}}$) also rise. This demonstrates the indirect effect of the number density $n_{\text{th}}$ on the free susceptibility $\chi_{f}$, as increased thermal fluctuations enhance the latter.
In contrast, the inverse dependence on $N^2$ shows a quantum-magnetic effect. As $N$ increases with magnetic flux, the larger degeneracy distributes particles among more equivalent states, thereby reducing the free susceptibility. Finally, the rotational dynamics of the system are expressed by the moment of inertia $I$. A larger moment of inertia signifies a stronger rotational response, which increases the pressure sensitivity to external parameters, such as the magnetic field or rotation, through fluctuations of the rotational degrees of freedom.
\par
As concerns $\kappa_{T}$ from \eqref{C16}, its proportionality to the moment of inertia indicates that a stronger rotational response enhances the system's sensitivity to changes in pressure or chemical potential, because a larger moment of inertia facilitates particle redistribution. Additionally, its proportionality to $1/N^{2}$ suggests that Landau-level degeneracy reduces the compressibility. This is because high degeneracy spreads particles across more equivalent quantum states and weakens the density's response to thermodynamic variations. Furthermore, the inverse dependence on $n_{\text{th}}$ implies that denser systems are less responsive to external parameters, resulting in lower compressibility.
\par
\begin{table*}[ht]
    \centering
    \caption{The range of parameters $t,b$, and $v$ in two phenomenological cases QGP and NS. }
    \label{tab1}
    \vspace{1ex}
        \begin{tabular}{lcccccccccccc}
            \hline\hline
\\
&\quad\quad&$t=T/m_{\pi}$&\qquad\qquad&$b=eB_{0}/m_{\pi}^{2}$&\qquad\qquad&$R$&\qquad\qquad&$v=R\Omega$&\qquad\qquad&$N$&\qquad\qquad&$\rho=\rho_{0}/m_{\pi}^{3}$\\ \hline \\
Case 1: QGP&\quad&$0.1\leq t\leq 1.5$&\qquad&$1\leq b\leq 10$ \cite{skokov2009}&\qquad &$10$ fm \cite{zahed2016}&\qquad &$0\leq v\leq 0.5$ \cite{voloshin2016}&\qquad&$24$ \cite{zahed2016}&\qquad&$0.28$ \cite{sahoo2025}\\
Case 2: NS&\quad&$10^{-4}\leq t\leq 0.1$&\qquad&$0.003\leq b\leq 0.03$ \cite{dexheimer2021}&\qquad &$10$ km \cite{hooman2023}&\qquad &$0\leq v\leq 0.02$ \cite{fukushima2015}&\qquad&$10^{4}$ \cite{fukushima2015}&\qquad&$0.012$ \cite{rojas2025}\\ \hline

            \hline\hline
        \end{tabular}
\end{table*}
\begin{figure}[h]
\includegraphics[width=8cm, height=6cm]{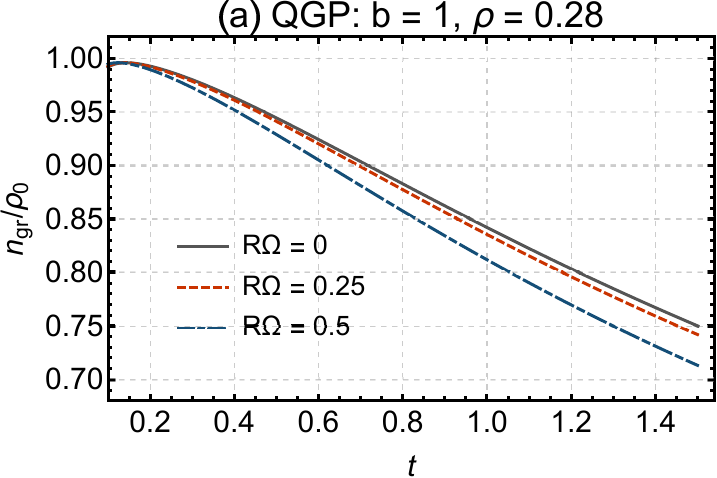}\vspace{0.5cm}
\includegraphics[width=8cm, height=6cm]{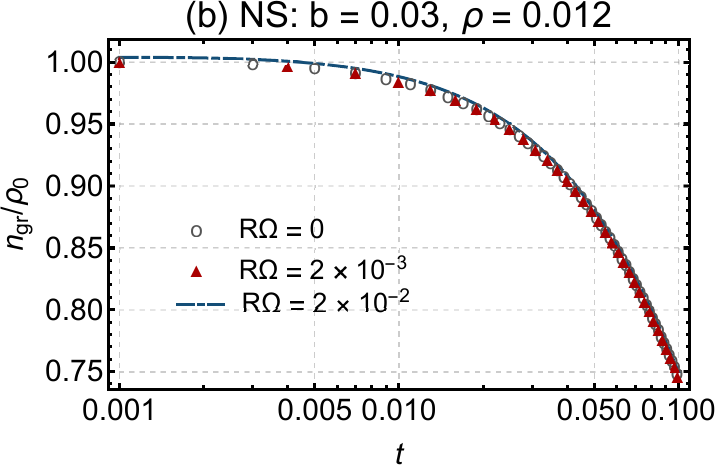}
\caption{The temperature ($t$) dependence of the ratio $n_{\text{gr}}/\rho_{0}$ is plotted for the QGP (panel a) and the NS (panel b) cases, considering various linear velocities $R\Omega$. In the temperature regimes relevant for QGP and NS cases, $n_{\text{gr}}/\rho_{0}$ is always smaller than $1$. In the QGP case, this ratio decreases as $t$ and $R\Omega$ increase. In the NS case, however,  $R\Omega$ does not significantly change the results. }\label{fig3}
\end{figure}
\begin{figure}[!h]
\includegraphics[width=8cm, height=6cm]{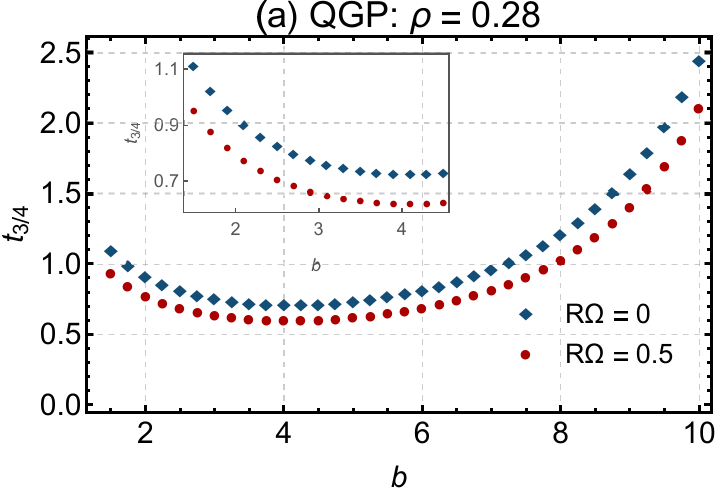}\vspace{0.5cm}
\includegraphics[width=8cm, height=6cm]{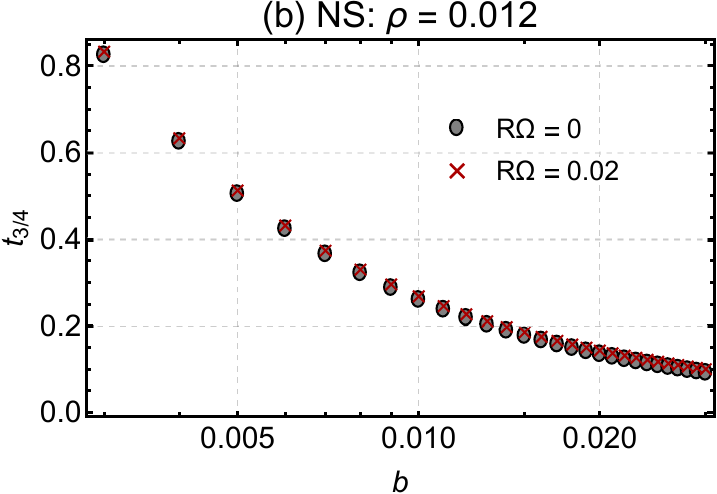}
\caption{The dependence of $t_{3/4}$ on the magnetic field ($b$) is plotted for the QGP (panel a) and the NS  (panel b) cases, considering various linear velocities $R\Omega$. The subdiagram in panel a shows how
$t_{3/4}$ varies with $b$-dependence for weak magnetic fields. The results of the QGP case indicate an effect similar to (inverse) magnetorotational catalysis of chiral symmetry breaking for (weak) strong magnetic fields. For the NS case, $t_{3/4}$ decreases with increasing $b$, indicating an inverse magnetorotational effect.}\label{fig4}
\end{figure}
\begin{figure*}[t]
\includegraphics[width=\textwidth]{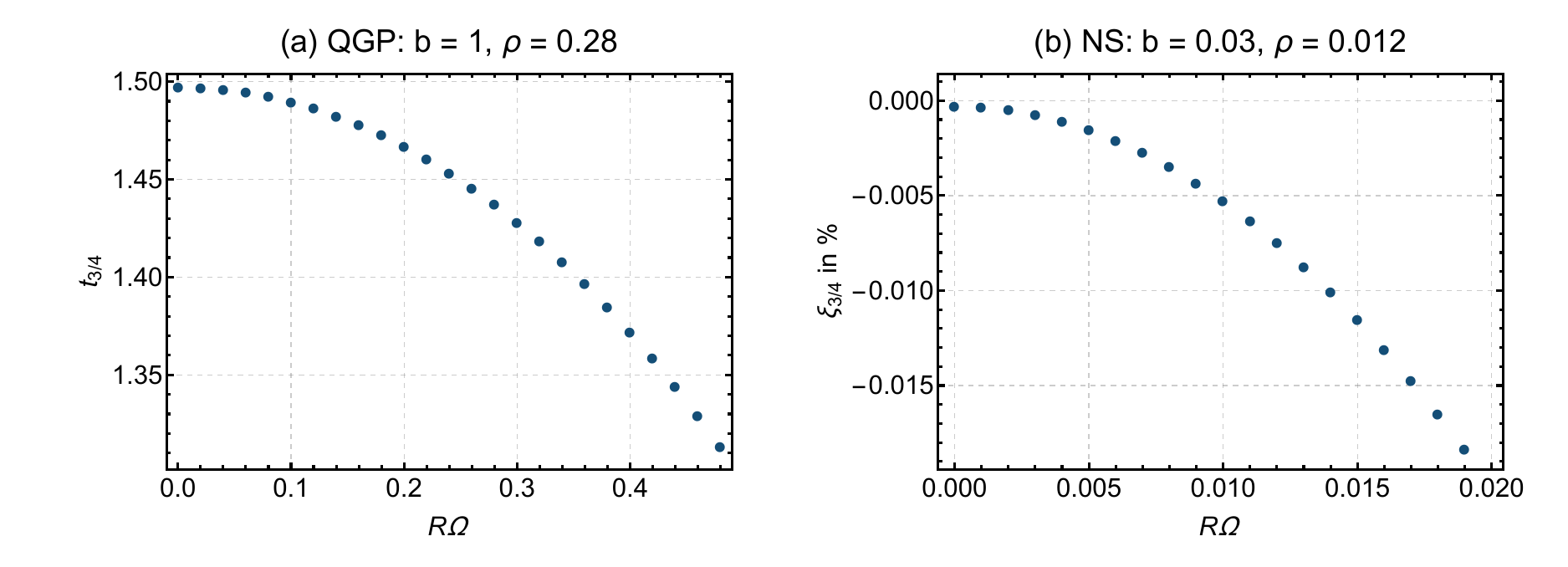}
\caption{The $R\Omega$ dependence of $t_{3/4}$ and $\xi_{3/4}$ is plotted for the QGP case (panel a) and the NS case (panel b). The results indicate an effect similar to inverse magnetorotational catalysis of chiral symmetry breaking for weak magnetic fields.}\label{fig5}
\end{figure*}
\begin{figure*}[hbt]
\includegraphics[width=\textwidth,height=15cm]{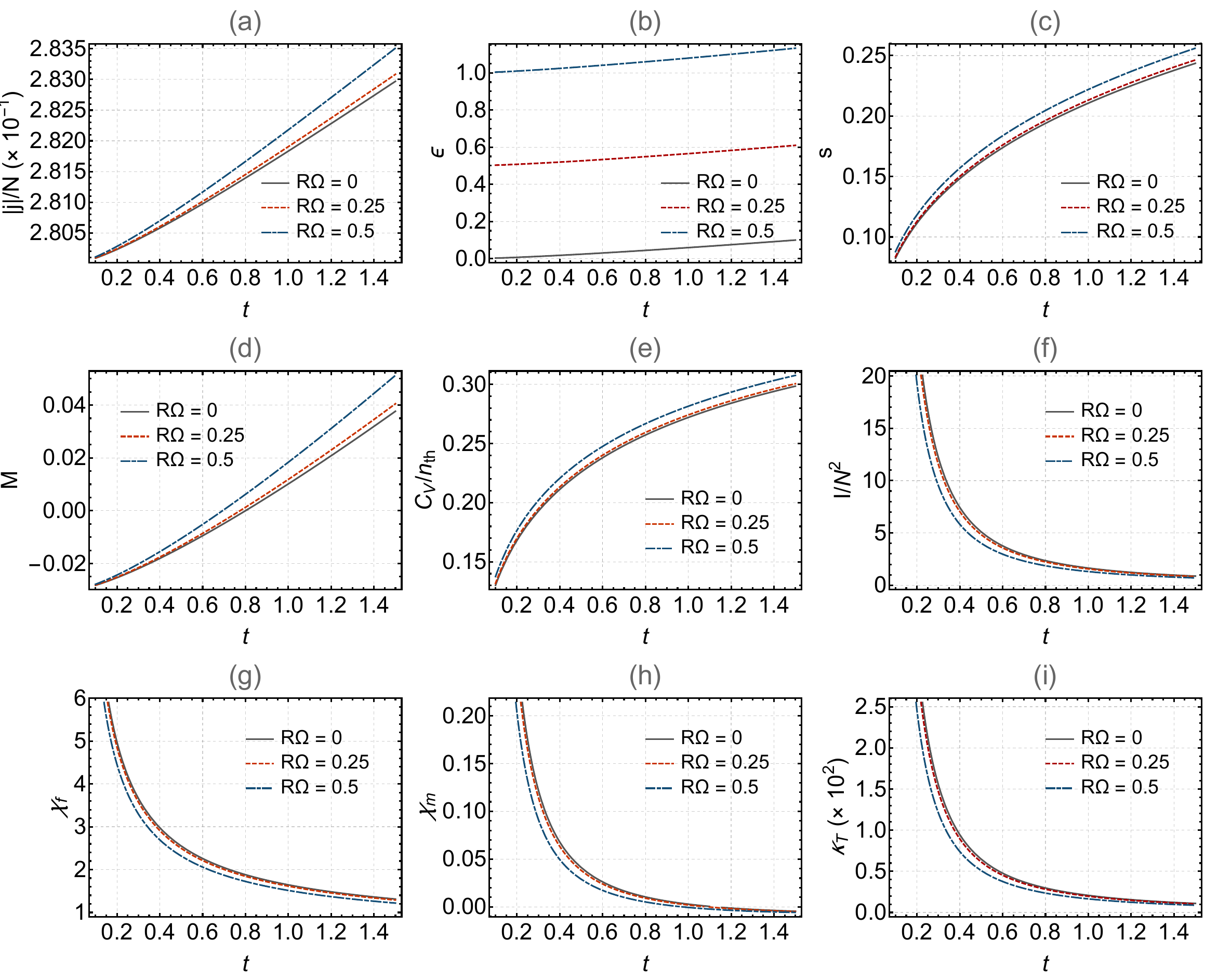}
\caption{The $t$ dependence of various thermodynamical quantities are plotted for the QGP case and various linear velocities. Here, $b=1$ and $\rho=0.28$. The results indicate that $|j|/N, \epsilon,s$, $M$, and $C_{V}/n_{\text{th}}$ increase with increasing $t$ and $R\Omega$. Other response functions, $I/N^{2}, \chi_{f},\chi_{m}$, and $\kappa_{T}$ decrease with increasing $t$ and $R\Omega$. In general, the results indicate that the QGP plasma behaves as a fluid. The behavior of $M$ is elaborated in Figs. \ref{fig7} and \ref{fig8}.}\label{fig6}
\end{figure*}
\begin{figure*}[t]
\includegraphics[width=8cm, height=6cm]{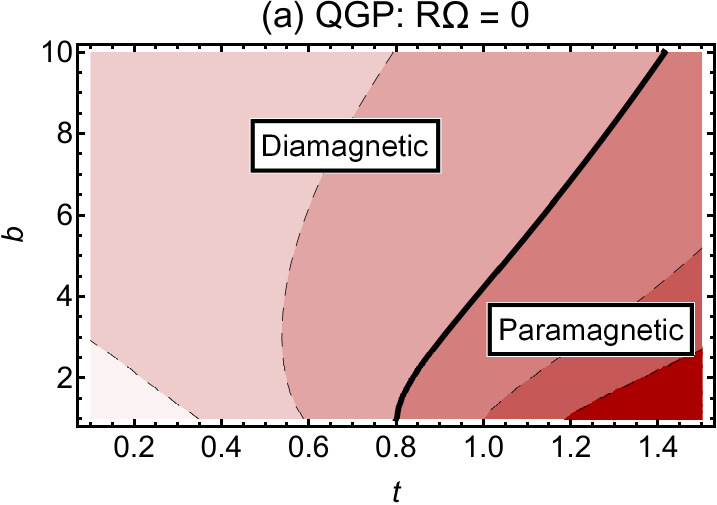}\hspace{0.1cm}
\includegraphics[width=8cm, height=6cm]{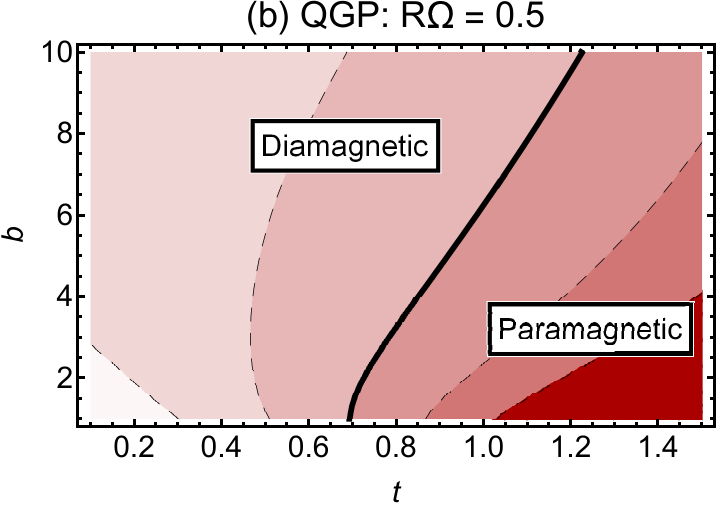}\hspace{0.1cm}
\par\vspace{0.3cm}\noindent
\includegraphics[width=8cm, height=6cm]{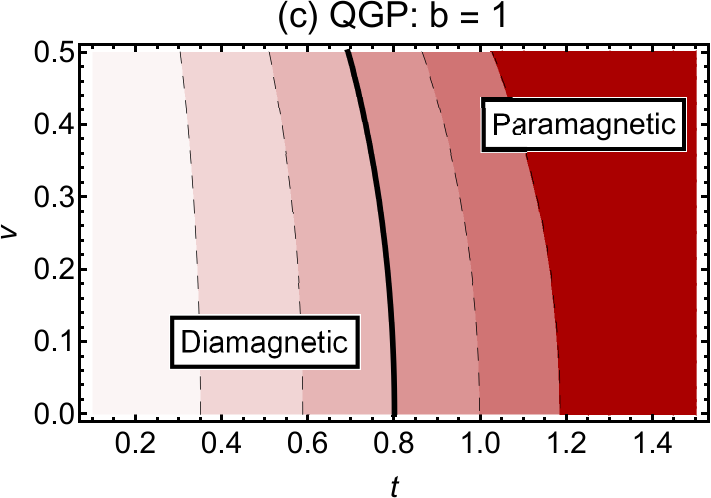}\hspace{0.1cm}
\includegraphics[width=8cm, height=6cm]{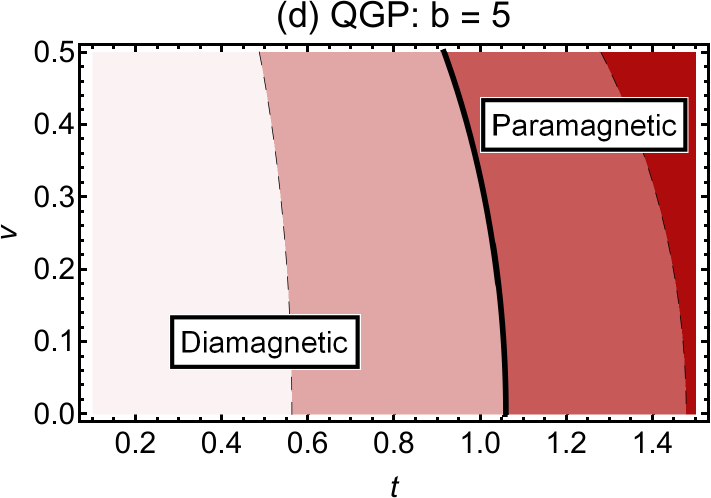}
\begin{center}
\includegraphics[width=10cm, height=0.8cm]{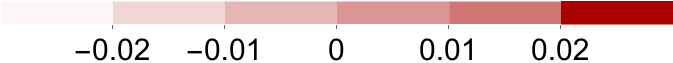}
\end{center}
\caption{In panel (a) and (b) contour plots $b$ versus $t$ for the magnetization $M$ of a QGP medium for $R\Omega=0$ (panel a) and $R\Omega=0.5$ (panel b) are displayed. In panel (c) and (d) contour plots $v=R\Omega$ versus $t$ for the magnetization $M$ of a QGP medium for $b=1$ (panel c) and $b=5$ (panel d) are plotted. Black solid lines indicate $M=0$ line. The results indicate that the diamagnetic phase space is enlarged by increasing $b$ and decreasing $R\Omega$. }\label{fig7}
\end{figure*}
\begin{figure}[h]
\includegraphics[width=8cm, height=6cm]{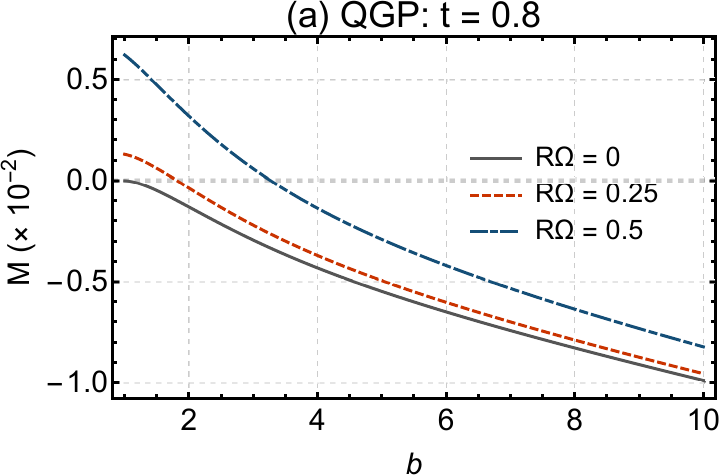}\vspace{0.5cm}
\includegraphics[width=8cm, height=6cm]{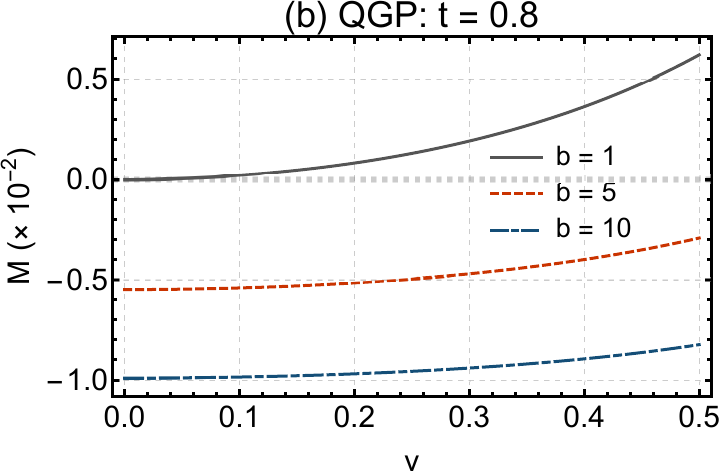}
\caption{Panel a: The $b$ dependence of $M$ for the QGP is plotted at a fixed temperature $t=0.8$ and for $R\Omega=0,0.25,0.5$. Panel b:
 The $v=R\Omega$ dependence of $M$ for the QGP medium is plotted at a fixed temperature $t=0.8$ and for $b=1,5,10$. The results indicate that the magnetic field suppresses the magnetization, while the rotation enhances it. The rotating Bose gas is paramagnetic in the weak $b$ field and large $R\Omega$ regime. }\label{fig8}
\end{figure}
\begin{figure*}[t]
\includegraphics[width=\textwidth]{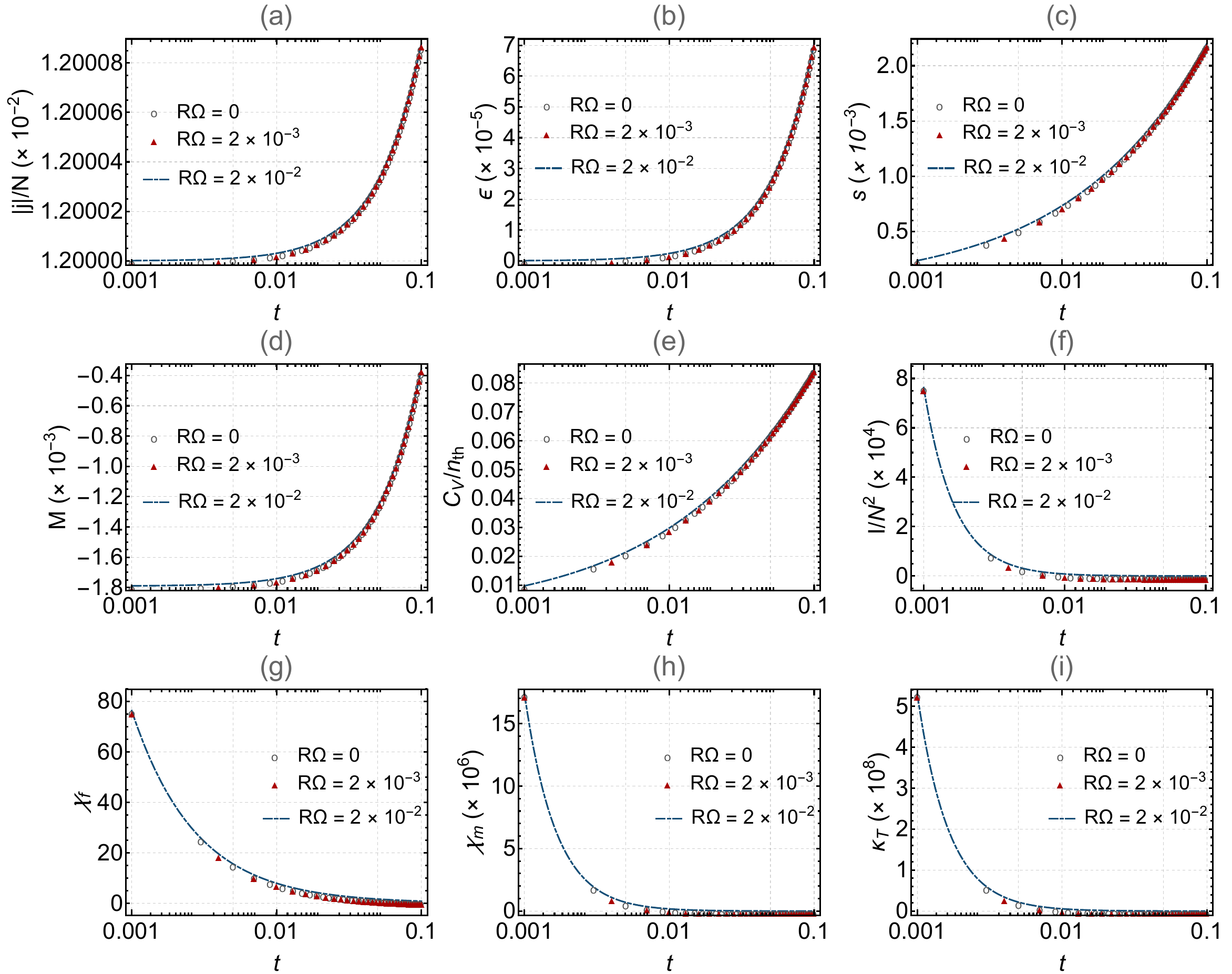}
\caption{The $t$ dependence of various thermodynamical quantities are plotted for the NS case and various linear velocities. Here, $b=0.03$ and $\rho=0.012$. The results indicate that $|j|/N, \epsilon,s$, $M$, and $C_{V}/n_{\text{th}}$ increase with increasing $t$. Other response functions, $I/N^{2}, \chi_{f},\chi_{m}$, and $\kappa_{T}$ decrease with increasing $t$. The results are not significantly altered by $R\Omega$. In contrast to the QGP case, $M$ is always negative indicating that the NS medium is diamagnetic in the LLL approximation. }\label{fig9}
\end{figure*}
\begin{figure*}[t]
\includegraphics[width=8cm, height=6cm]{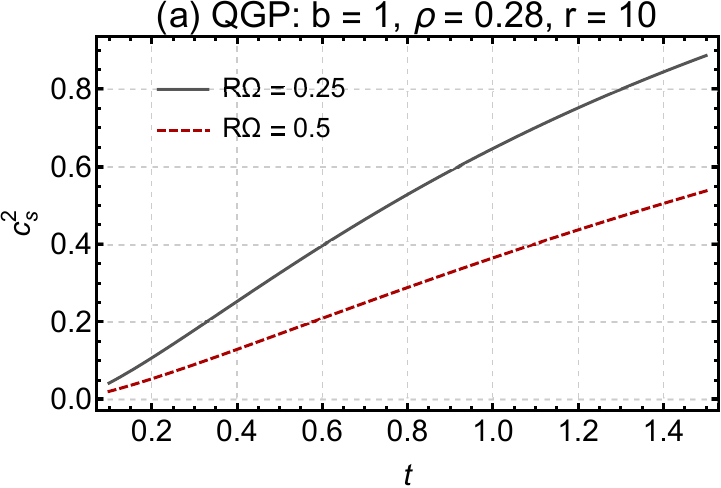}\hspace{0.2cm}
\includegraphics[width=8cm, height=6cm]{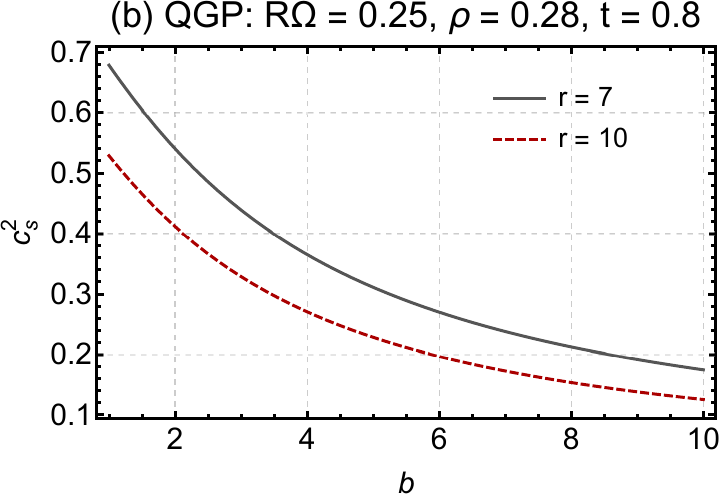}
\caption{Panel a: The $t$ dependence of $c_{s}^{2}$ is plotted for $b=1, \rho=0.28, r=10$, and $R\Omega=0.25,0.5$ for the QGP case.  Panel b: The $b$ dependence of $c_{s}^{2}$ is plotted for $R\Omega=0.25, \rho=0.28, t=0.8$, and $r=7,10$ for the QGP case. }\label{fig10}
\end{figure*}
\begin{figure*}[t]
\includegraphics[width=8cm, height=6cm]{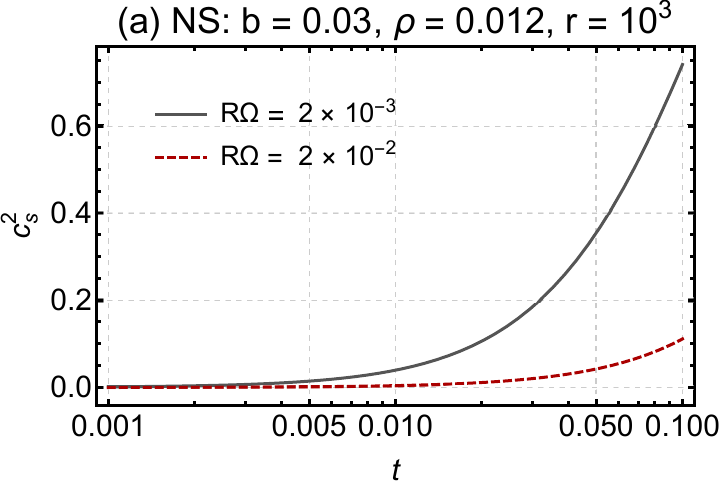}\hspace{0.2cm}
\includegraphics[width=8cm, height=6cm]{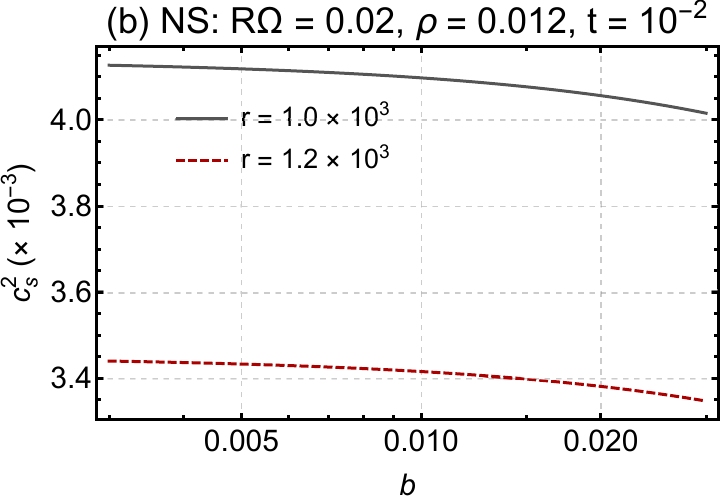}
\caption{Panel a: The $t$ dependence of $c_{s}^{2}$ is plotted for $b=0.03, \rho=0.012, r=10^{3}$, and $R\Omega=0.002,0.02$ for the NS case.  Panel b: The $b$ dependence of $c_{s}^{2}$ is plotted for $R\Omega=0.02, \rho=0.012, t=10^{-2}$, and $r=10^{3}, 1.2\times 10^{3}$.}\label{fig11}
\end{figure*}
\begin{figure*}[t]
\includegraphics[width=\textwidth]{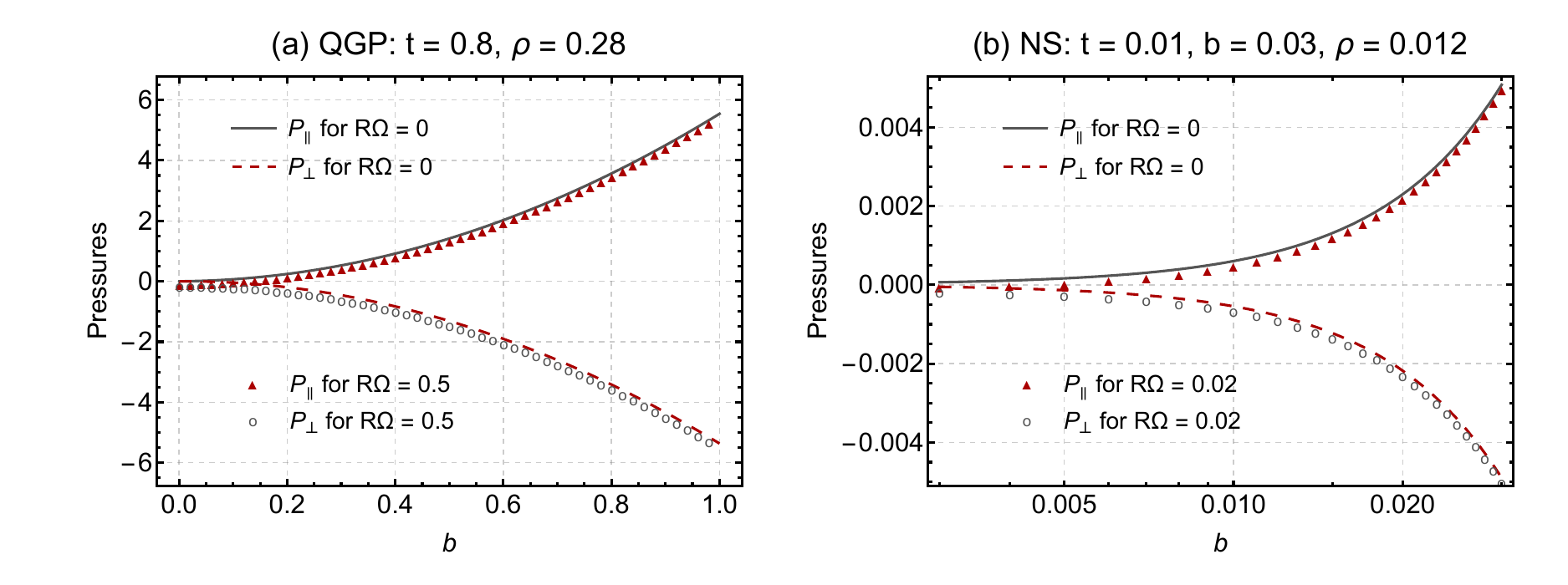}
\caption{Panel a: The $b$ dependence of $P_{\|}$ and $P_{\perp}$ is plotted for the QGP case (panel a) and the NS case (panel b) and various linear velocities $R\Omega$. We observe that $R\Omega$, arising from relativistic correction to temperature does not significantly alter the results, neither in the QGP nor in the NS medium.}\label{fig12}
\end{figure*}
\begin{figure*}[t]
\includegraphics[width=8cm, height=6cm]{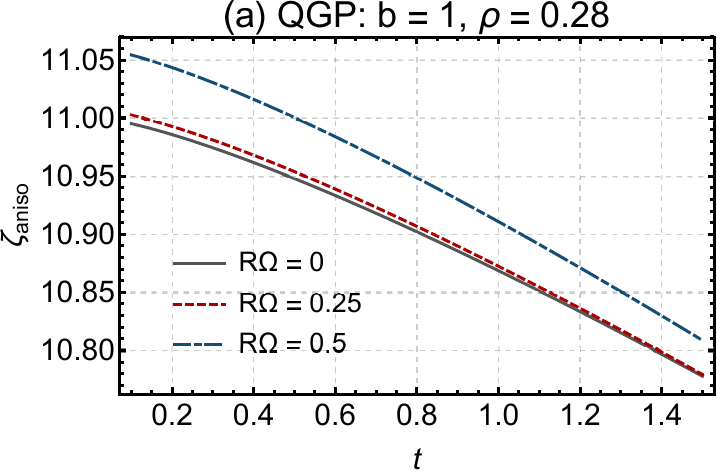}\hspace{0.2cm}
\includegraphics[width=8cm, height=6cm]{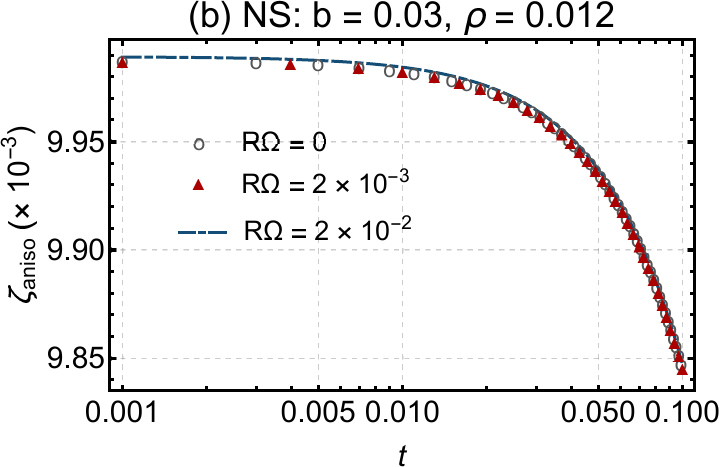}
\caption{The $t$ dependence of $\zeta_{\text{aniso}}$ is plotted for the QGP case (panel a) and NS case (panel b) for various linear velocities $R\Omega$. According to these results, in the QGP medium $\xi_{\text{aniso}}$ decreases with increasing $t$ and decreasing $R\Omega$. In the NS medium, however, $R\Omega$ does not significantly alter the results. }\label{fig13}
\end{figure*}
Another important quantity related to the thermodynamic properties of the magnetized and rotating gas is the  sound velocity $c_{s}$, whose squared is defined by
\begin{eqnarray}\label{C22}
c_{s}^{2}\equiv\left(\frac{dP}{d\epsilon}\right)_{s/n,B,\Omega}.
\end{eqnarray}
To compute this expression, we reformulate it using the Gibbs-Duhem and the energy relations
\begin{eqnarray}\label{C23}
dP&=&sdT+nd\mu+jd\Omega+MdB,\nonumber\\
d\epsilon&=&Tds+\mu dn-jd\Omega-MdB,
\end{eqnarray}
and assuming constant $s/n, B$, as well as $\Omega$. We arrive at
\begin{eqnarray}\label{C24}
c_{s}^{2}=\frac{\frac{dP}{dT}}{\frac{d\epsilon}{dT}}\bigg|_{s/n,B,\Omega}=\frac{\frac{\partial P}{\partial T}+\frac{\partial P}{\partial\mu}\frac{d\mu}{dT}}{\frac{\partial \epsilon}{\partial T}+\frac{\partial \epsilon}{\partial\mu}\frac{d\mu}{dT}}\bigg|_{s/n,B,\Omega}.
\end{eqnarray}
Using $d(s/n)/dT=0$, we obtain
\begin{eqnarray}\label{C25}
\frac{d\mu}{dT}=\frac{n_{\text{th}}\frac{\partial s}{\partial T}-s\frac{\partial n_{\text{th}}}{\partial T}}{s\frac{\partial n_{\text{th}}}{\partial \mu}-n_{\text{th}}\frac{\partial s}{\partial \mu}}.
\end{eqnarray}
Additionally, utilizing $P=-\Phi_{T}^{\text{LLL}}$ and the definitions of $n_{\text{th}}$ from \eqref{B3}, $s$ from \eqref{C6}, $\chi_{f}$ from \ref{C2} as well as relations from \eqref{appB1}, we also have
\begin{eqnarray*}
\frac{\partial P}{\partial T}&=&s, \qquad \frac{\partial P}{\partial \mu}=n,\qquad \frac{\partial n_{\text{th}}}{\partial \mu}=\frac{\chi_{f}}{T},\nonumber\\
\frac{\partial s}{\partial \mu}&=&\frac{n_{\text{th}}}{2T}-\frac{\chi_{f}}{T^{2}}\left(\mu_{B}-N\Omega\right),\nonumber\\
\frac{\partial n_{\text{th}}}{\partial T}&=&\frac{n_{\text{th}}}{2T}-\frac{\chi_{f}}{T^{2}}\left(\mu_{B}-N\Omega\right),
\end{eqnarray*}
\begin{eqnarray}\label{C26}
\frac{\partial s}{\partial T}&=&\frac{s}{2T}-\frac{n_{\text{th}}}{2T^{2}}\left(\mu_{B}-N\Omega\right)+\frac{\chi_{f}}{T^{3}}\left(\mu_{B}-N\Omega\right)^{2}, \nonumber\\
\frac{\partial \epsilon}{\partial T}&=&\frac{s}{2}+\frac{n_{\text{th}}}{2T}N\Omega-\frac{\chi_{f}}{T^{2}}N\Omega\left(\mu_{B}-N\Omega\right),\nonumber\\
\frac{\partial\epsilon}{\partial\mu}&=&\frac{n_{\text{th}}}{2}+\frac{\chi_{f}}{T}N\Omega.
\end{eqnarray}
Plugging these expressions into \eqref{C25} and the resulting expression into \eqref{C24}, it is possible to explore the temperature dependence of the speed of sound [see Sec. \ref{sec5}] for fixed $eB_{0}$ and $\Omega$ as well as the number density $\rho_{0}$.
\par
Apart from the thermodynamic quantities mentioned above, we will determine the pressure anisotropy factor $\zeta_{\text{aniso}}$  in Sec. \ref{sec5}. This factor arises from the presence of a constant magnetic field $\bs{B}$ and angular-momentum $\bs{\Omega}$. It is derived from the energy-momentum tensor in the corotating frame $T^{\mu\nu}$. As it is described in \cite{jafari2023, kiamari2025}, $T^{\mu\nu}$ is given by
\begin{eqnarray}\label{C27}
T^{\mu\nu}=\left(
\begin{array}{cccc}
\epsilon&y\Omega\epsilon&-x\Omega\epsilon&0\\
y\Omega\epsilon&\mathcal{P}_{\perp}+y^{2}\Omega^{2}\epsilon&-xy\Omega^{2}\epsilon&0\\
-x\Omega\epsilon&-xy\Omega^{2}\epsilon&\mathcal{P}_{\perp}+x^{2}\Omega^{2}\epsilon&0\\
0&0&0&P_{\|}
\end{array}
\right), \nonumber\\
\end{eqnarray}
where $\mathcal{P}_{\perp}\equiv P-MB+B^{2}/2$ and $P_{\|}\equiv P-B^{2}/2$ are the perpendicular and parallel pressures in the absence of rotation \cite{ferrer2010, fayazbakhsh2014}. We note that $T^{\mu\nu}$ arises from $T^{\mu\nu}=e^{\mu}_{~a}e^{\nu}_{~b}T^{ab}$ with the vierbeins $e^{\mu}_{~a}$ from \eqref{A3} and $T^{ab}$ the energy momentum tensor in a nonrotating inertial frame in the presence of magnetic field \cite{tabatabaee2019}. To define $\zeta_{\text{aniso}}$, we first define $P_{\perp}$ by
\begin{eqnarray}\label{C28}
P_{\perp}\equiv \frac{1}{2}\left(T^{xx}+T^{yy}\right)=\mathcal{P}_{\perp}+\frac{1}{2}v^{2}\epsilon,
\end{eqnarray}
where $R^{2}=x^{2}+y^{2}$ is used. The anisotropy factor is then given by
\begin{eqnarray}\label{C29}
\zeta_{\text{aniso}}\equiv P_{\perp}-P_{\|}.
\end{eqnarray}
\section{Numerical Results}\label{sec5}
\setcounter{equation}{0}
In Sec. \ref{sec2B}, we determined the thermodynamic potential of a magnetized Bose gas under rigid rotation. In Sec. \ref{sec3}, we considered the LLL contribution to this potential and analyzed the weak scenario of BEC in the presence of $\Omega$ and $B$. In particular, we derived the magnetorotational fugacity of this medium by assuming a constant number density and applying an appropriate regularization scheme. The analytical expression for $\Phi_{T}^{\text{LLL}}$ is utilized in Sec. \ref{sec4} to derive analytical expressions for thermodynamic quantities, including  $j,\epsilon, s, M,C_{V},I, \chi_{f}, \chi_{m}$, and $\kappa_{T}$.
\par
In this section, we first numerically determine the $T$ dependence of $z_{B,\Omega}$ within the weak scenario of BEC. We then explore the $T$ dependence of the aforementioned thermodynamic quantities. For simplicity, we introduce dimensionless variables
\begin{eqnarray}\label{D1}
t\equiv T/m_{\pi}\quad b\equiv eB_{0}/m_{\pi}^{2}, \quad \rho\equiv \rho_{0}/m_{\pi}^{3},
\end{eqnarray}
with the pion mass $m_{\pi}=0.140$ GeV, and rescale all thermodynamic quantities with $m_{\pi}$ (see Table \ref{tab1}). Additionally, we define the dimensionless linear velocity $v=R\Omega$. To determine the speed of sound, we introduce a new variable $r\equiv \bar{R}m_{\pi}$. Here, we replace $N$ with $N=br^{2}/2$ [see \eqref{A20}].
To fix free parameters $T$, $b$, as well as the number density $\rho$, linear velocity $v$, and the density of state $N$, we consider two phenomenological cases. In Case 1, we focus on the physics of QGP characterized by relatively large $\rho,T,\Omega$, and strong $B$. In Case 2, we examine the physics of NS matter, where the values of $\rho, T, \Omega$, and $B$ are smaller. The parameters for these two cases are summarized in Table \ref{tab1}. Using
\begin{eqnarray}\label{D2}
1~\mbox{eV}&\sim& 10^{4}~K,\nonumber\\
 1~m_{\pi}&\sim& 0.7~\text{fm}^{-1},\nonumber\\
 1~\text{GeV}&\sim& 1.52\times 10^{24}~\text{Hz},\nonumber\\
eB_{0}=1~\text{GeV}^{2}&\sim& B=1.7\times 10^{20}~G.
\end{eqnarray}
it is possible to convert the values of $t,b$, and $\Omega$ from Table \ref{tab1} into Kelvin (K), Gau\ss~(G), and Hertz (Hz). In Figs. \ref{fig1}(a)-\ref{fig3}(a) and \ref{fig6}, \ref{fig10}(a), \ref{fig11}(a), and \ref{fig11}(c), we set $b=1$ \cite{skokov2009, zahed2016} and $\rho=0.28$ \cite{sahoo2025} for the QGP case. In Figs. \ref{fig1}(b)-\ref{fig3}(b) and \ref{fig9}, \ref{fig10}(b), \ref{fig11}(b), and \ref{fig11}(d), we choose $b=0.03$ \cite{dexheimer2021} and $\rho=0.012$ \cite{rojas2025} for the NS case.
\par
In Fig. \ref{fig1}, we plot the $t$ dependence of $z_{B,\Omega}$ from \eqref{B15} for two cases: the QGP case [Fig. \ref{fig1}(a)] and the NS case [Fig. \ref{fig1}(b)]. As mentioned in Sec. \ref{sec3}, the dependence of $\Omega$ from the term $N\Omega$ in $z_{B,\Omega}$ cancels out due to the regularization method we used to derive $z_{B,\Omega}$ from \eqref{B15}. We reintroduce the $\Omega$-dependence, we consider the relativistic correction of the temperature. In the case of QGP, with $R=10$ fm, the values $R\Omega=0.25$ and $R\Omega=0.5$ shown in Fig. \ref{fig1}(a) correspond to $\Omega\sim 7.5\times 10^{21}$ Hz and $\Omega\sim 1.5\times 10^{22}$ Hz, respectively. These extremely large angular velocities are thought to be generated in the early stages of heavy-ion collisions \cite{voloshin2016}. In contrast, for the NS case, using $R=10$ km \cite{hooman2023}, the values of $R\Omega=2\times 10^{-3}$ and $R\Omega=2\times 10^{-2}$ in Fig. \ref{fig1}(b) correspond to approximately $\Omega\sim 60.86$ Hz and $\Omega\sim 608.6$ Hz, respectively. In \cite{fukushima2015, hooman2023}, it is estimated that the angular velocity of neutron stars is around $\Omega \sim \mathcal{O}(10^{3})$. As demonstrated in Fig. \ref{fig1}, while relativistic corrections affect $z_{B,\Omega}$ in the QGP case, the magnetorotational fugacity in the NS case remains independent of $v=R\Omega$. In both cases $z_{B,\Omega}<1$, which is one of the necessary conditions for the weak scenario of BEC.
\par
In Fig. \ref{fig2}, the $t$ dependence of the pressure $P=-\Phi_{T}^{\text{LLL}}$ is presented. As in previous sections, we focus solely on the LLL contribution to the thermodynamic potential. The expression for $\Phi_{T}^{\text{LLL}}$ from \eqref{A35} depends explicitly on $z_{B,\Omega}$. We use the data for $z_{B,\Omega}$, derived from \eqref{B15} and displayed in Fig. \ref{fig1}, to calculate $P$.
As expected, in both in the QGP [Fig. \ref{fig2}(a)] and NS [Fig. \ref{fig2}(b)] cases, the pressure increases monotonically with rising temperature. In the QGP case, at a fixed temperature, $P$ increases with an increase in $v=R\Omega$. In contrast, the relativistic corrections in the NS case do not significantly affect the results.
\par
Plugging $z_{B,\Omega}$ from \eqref{B15} into \eqref{B10}, we can determine the number density $n_{\text{gr}}$. For the case of QGP (NS), we fix the cutoff $p_{0}$ by requiring that for $b=1$ ($b=0.03$) and for $\rho=0.28$ ($\rho=0.012$), $n_{\text{gr}}(t=1.5)=0.75 \rho$ ($n_{\text{gr}}(t=0.1)=0.75 \rho$). In the QGP case, $p_{0}\sim 0.997 m_{\pi}$, while in the NS case $p_{0}\sim 0.073 m_{\pi}$. It is noteworthy that the main physical outcomes are independent of the cutoff $p_{0}$.
\par
Figure \ref{fig3} illustrates the $t$ dependence of $n_{\text{gr}}/\rho$ for both the QGP case [Fig. \ref{fig3}(a)] and the NS case [\ref{fig3}(b)]. According to our results, $n_{\text{gr}}$ decreases as the temperature increases. In both cases, we substitute $t$ with $\Gamma(v)t$ to account for relativistic corrections to the temperature. For the QGP case, we set $v=0,0.25, 0.5$, and for the NS case, we choose $v=0, 0.002, 0.02$.
\par
To explore the dependencies of $n_{\text{gr}}$ on $b$ and $\Omega$ at a fixed temperature, we determine the temperature at which the condensate $n_{\text{gr}}$ decreases to $75$ percent of its original value at $t=0$. We refer to this temperature as $t_{3/4}$.  Specifically, we determine $t_{3/4}$ from
\begin{eqnarray}\label{D3}
n_{\text{gr}}(t_{3/4}(\bar{b},\bar{v},\rho), \bar{b}, \bar{v})=0.75 \rho,
\end{eqnarray}
where $\bar{b}$ and $\bar{v}$ represent a fixed magnetic field and a given linear velocity. In Fig. \ref{fig4}, the $b$ dependence of $t_{3/4}$ is demonstrated for two cases: QGP and NS phenomenology. For the QGP case, we set $\rho=0.28$ and $R\Omega=0,0.5$, and determine $t_{3/4}$ for $1\leq b\leq 10$ [see Fig. \ref{fig4}(a)]. For the NS case, we choose $\rho=0.012$ and $R\Omega=0,0.02$, calculating $t_{3/4}$ for $0.003\leq b\leq 0.03$ [see Fig. \ref{fig4}(b)]. In Fig. \ref{fig4}(a), $t_{3/4}$ initially decreases for $1<b<4.5$ and then increases for $4.5<b<10$. The subdiagram in Fig. \ref{fig4}(a) scrutinizes the $b$ dependence of $t_{3/4}$ within the interval $b\in[1,4.5]$ while keeping the same free parameters $\rho$ and $v$. According to these results, in the context of QGP phenomenology, the formation of BE condensates is suppressed for $b<4.5$ and enhanced for larger $b>4.5$.
\par
This behavior of $t_{3/4}$ with respect to $b$ resembles the phenomena of (inverse) magnetic catalysis, where a magnetic field enhances (suppresses) the formation of condensates resulting from spontaneous symmetry breaking.  It is important to note that the magnetic field created in the early stages of HICs at RHIC and LHC is estimated to be $eB\sim 1.5 m_{\pi}^{2}$ and $eB\sim 15 m_{\pi}^{2}$. Although the strength of the $B$ field decreases rapidly as temperature (or time) decreases, for $R=10$ fm, the field remains strong enough that the discussion of magnetic catalysis or inverse magnetic catalysis is quite relevant to BEC physics in QGP. We also observe that for $R\Omega=0.5$, the value of $t_{3/4}$ is lower than for $R\Omega=0$. This is expected, as the rotation suppresses the formation of condensates in a process similar to an inverse magnetorotational one \cite{sadooghi2021} [see also Fig. \ref{fig5}]. Using the same method, we calculate $t_{3/4}$ for NS phenomenology with $\rho=0.012$, $R\Omega=0,0.02$, and $0.003<b<0.03$. The results are presented in Fig. \ref{fig4}(b). According to these findings, $t_{3/4}$ decreases as $b$ increases. This behavior aligns with the concept of inverse magnetic catalysis in the presence of a weak magnetic field and in the absence of rotation \cite{ayala2012}. The phenomenon of magnetic catalysis for strong magnetic fields is also discussed in \cite{ayala2012}. Additionally, the data from Fig. \ref{fig4}(b) indicate that the relativistic correction of the temperature does not significantly alter the results.
\par
In Fig. \ref{fig5}(a), the $R\Omega$ dependence of $t_{3/4}$ is plotted for fixed $b=1$ and $\rho=0.28$ in the QGP case. As expected from the results presented in Fig. \ref{fig4}(a), $t_{3/4}$ decreases with increasing $R\Omega$. This behavior is analogous to the inverse magnetorotational effect observed in the spontaneous chiral symmetry breaking, as explored in \cite{sadooghi2021, hooman2023}. These findings indicate that rotation suppresses the formation of BE condensates. In the case of NS phenomenology, for a weak magnetic field $b=0.03$ and a fixed $\rho=0.012$, the effect of rotation is minimal, as is also expected based on Fig. \ref{fig4}(b). To demonstrate this point, we introduce a new quantity, $\xi_{3/4}$, defined by
\begin{eqnarray}\label{D4}
\xi_{3/4}\equiv \frac{t_{3/4}(\bar{b},\bar{v},\rho)-t_{3/4}(\bar{b},0,\rho)}{t_{3/4}(\bar{b},0,\rho)}\times 100.
\end{eqnarray}
In Fig. \ref{fig5}(b), the dependence of $\xi_{3/4}$ on $v=R\Omega$ is plotted for $b=0.03$ and $\rho=0.012$ in the context of NS phenomenology. The decrease of $\xi_{3/4}$ with increasing $R\Omega$ indicates an inverse magnetorotational catalysis effect in the formation of BE condensate, similar to the phenomenon observed in spontaneous chiral symmetry breaking \cite{sadooghi2021, hooman2023}.
\par
In Figs. \ref{fig6} and \ref{fig9}, the $t$ dependence of thermodynamic quantities such as $|j|/N, \epsilon, s,M, C_{V}/n_{\text{th}}, I/N^{2}, \chi_{f}, \chi_{m}$, and $\kappa_{T}$ from \eqref{C4}-\eqref{C7}, \eqref{C10}, \eqref{C12}, \eqref{C13}, \eqref{C15}, and \eqref{C17} is plotted for a) the QGP [Figs. \ref{fig6}(a)-\ref{fig6}(i)] and b) NS [Figs. \ref{fig9}(a)-\ref{fig9}(i)] phenomenologies. These quantities are expressed in terms of $\mbox{Li}_{\nu}(z_{B,\Omega})$. Plugging $z_{B,\Omega}$, obtained from \eqref{B15}, into the $\mbox{Li}_{\nu}(z_{B,\Omega})$, we arrive at the $t$ dependence of the above thermodynamic quantities.
The results indicate that the ratio of angular-momentum density to the Landau degeneracy factor $|j|/N$ increases with temperature $t$. Since $|j|=N n$, as mentioned above, the $t$ dependence of the ratio $|j|/N$ simultaneously reflects the dependence of the number density $n$ on temperature. The latter increases as $t$ increases, as expected.
\par
Regarding the $R\Omega$ dependence, which arises from the relativistic correction to the temperature, it turns out that in the QGP case, for a fixed value of $t$, the ratio $|j|/N$ increases with increasing $R\Omega$. The trend holds true for $\epsilon$, $s$, $M$, and $C_{V}/n_{\text{th}}$ in Figs. \ref{fig6}(b)-\ref{fig6}(e) for the QGP case. However, in the NS case, the relativistic correction for $t$ does not significantly alter the results as shown in Figs. \ref{fig9}(a)-\ref{fig9}(i). In the both QGP and NS cases, the specific heats  $C_{V}$ do not exhibit any peaks and instead increase monotonically with rising temperature. This monotonic behavior is the final necessary condition that ensures the possibility of the weak scenario of BEC, as discussed in Sec. \ref{sec2}. Figures \ref{fig6}(f)- \ref{fig6}(i) (QGP case) and Figs. \ref{fig9}(f)-\ref{fig9}(i) (NS case) display the $t$ dependence of the response functions $I,\chi_{f}, \chi_{m}$ and $\kappa_{T}$. For both cases, all response functions decrease monotonically as the temperature increases.
\par
In Fig. \ref{fig6}(d), the $t$ dependence of the magnetization $M$ is presented for $b=1$ and for $R\Omega=0,0.25$, and $0.5$. At low temperatures, specifically for $t<0.8$ and $R\Omega=0$, we observe that the magnetization is negative. It reaches zero at a specific temperature $t_{*}\sim 0.8$ and becomes positive at $t>t_{*}$. As we increase $R\Omega$ to values greater than zero, the value of $t_{*}$ shifts. The data in Fig. \ref{fig6}(d) indicate that with an increase in $R\Omega$, the magnetization $M$ increases, while $t_{*}$ decreases. According to \cite{rojas2025}, a medium with $M<0$ is classified as diamagnetic, whereas a medium with $M>0$ is classified as paramagnetic. The results shown in Fig. \ref{fig6}(d) indicate that $R\Omega$ affects  the magnetic properties of the rotating medium.
\par
In Figs. \ref{fig7} and \ref{fig8}, we examine this phenomenon in the context of QGP. In Fig. \ref{fig8}(a), we present a contour plot showing the relationship of $b$ and $t$ for fixed values of $R\Omega=0$ and $0.5$. Figure \ref{fig7}(b) illustrates a contour plot of the linear velocity $v=R\Omega$ versus $t$ for two fixed magnetic field strengths, $b=1$ [Fig. \ref{fig7}(c)] and $b=5$ [Fig. \ref{fig7}(d)]. The regions where $M>0$ are labeled "paramagnetic", while regions with $M<0$ are labeled "diamagnetic". The solid black lines in these figures separate these regions and indicate the points where $M=0$. The results demonstrate that increasing $R\Omega$ enhances paramagnetism, whereas the phase space for $M>0$ contracts as $b$ increases. Conversely, the magnetic field tends to induce a diamagnetic response in the free Bose gas.
\par
In Fig. \ref{fig8}(a), we plot the $b$ dependence of $M$ on $b$ at a fixed temperature of $t=0.8$ for $R\Omega=0,0.25,0.5$. In Fig. \ref{fig8}(b), we illustrates how $M$ varies with $v$ at the same fixed temperature for $b=1,5,10$. The results in Fig. \ref{fig8}(a) indicate that the medium exhibits diamagnetism when $R\Omega=0$. As $b$ increases, the magnetization $M$ decreases. However, when $R\Omega>0$, $M$ becomes positive for $b<b_{*}$ and turns negative again when $b>b_{*}$. Our findings suggest that $b_{\star}$ increases as $R\Omega$ rises. Additionally, the dependence of $M$ on $v$ for $b=1,5,10$ in Fig. \ref{fig8}(b) indicates that $R\Omega$ positively influences the magnetization of the medium. For small $b$, it can even convert a diamagnetic free Bose gas into a paramagnetic one.
\par
We observe that for the parameters relevant to the NS case, the magnetization $M$ is always negative as shown in Fig. \ref{fig9}(d). Additionally, $R\Omega$ is insignificant enough that it does not affect the magnetic properties of the medium. It is important to emphasize that the conclusion drawn above, particularly regarding the effect of the magnetic field, is based on the approximations made in this paper. These results are consistent with findings presented in \cite{rojas2025}, which discuss the effect of magnetic fields on the thermodynamics of nonrotating neutron stars.
\par
In Sec. \ref{sec4}, we define the sound velocity $c_{s}$ of the medium [see \eqref{C22} and \eqref{C24}]. Figure \ref{fig10}(a) displays the $t$ dependence of $c_{s}^{2}$ for the QGP phenomenology. In this case, it is necessary to fix $r=\bar{R}m_{\pi}$, where $r=(2N/b)^{1/2}$ is associated with the Landau degeneracy factor $N$ from \eqref{A20}. This factor appears explicitly in the expressions of \eqref{C26}, which contribute to \eqref{C24}. In Fig. \ref{fig10}(a), the $t$ dependence of $c_{s}^{2}$ is plotted for $b=1, \rho=0.28$, and $r=10$. To investigate the effect of the relativistic corrections to $t$, we consider two cases: $R\Omega=0.25$ and $R\Omega=0.5$. We observe that $c_{s}^{2}$ increases as $t$ rises. Additionally, for a fixed value of $t$, $c_{s}^{2}$ decreases as $R\Omega$ increases. This behavior is related to the equation of state $\epsilon=\epsilon(p)$ from \eqref{C5}. In the case of $\Omega=0$ and $b\neq 0$, we find $\epsilon=P/2$. From \eqref{C22}, the speed of sound squared is given by $c_{s}^{2}=2$. Thus, in this scenario, the speed of sound equals the phase velocity defined as $\epsilon_{\text{ph}}\equiv P/\epsilon=\sqrt{2}$. However, when $\Omega\neq 0$, we expect $c_{s}^{2}<2$ and $c_{s}\neq c_{\text{ph}}$.
\par
In Fig. \ref{fig10}(b), the $b$ dependence of $c_{s}$ is demonstrated for fixed $R\Omega=0.25, \rho=0.28, t=0.8$, and $r$ set to $r=7$ and $r=10$. It is important to note that larger values for $r$ correspond to an increase in the degeneracy factor $N$. Our observations reveal that $c_{s}$ decreases as $b$ increases. Furthermore, for a fixed value of $b$, the speed of sound decreases with increasing $r$, or, equivalently, with increasing $N$.
The conclusions drawn from the parameters related to NS phenomenology remain consistent. In Fig. \ref{fig11}(a), the $t$ dependence of $c_{s}^{2}$ is plotted for $b=0.03, \rho=0.012, r=10^3$, and $R\Omega=0.002,0.02$. At low temperatures $t<0.005$, the effect of relativistic corrections to $t$ is negligible. However, at higher temperatures, $c_{s}^{2}$ decreases as $R\Omega$ increases. In Fig. \ref{fig11}(b), for $R\Omega=0.002, \rho=0.012, t=10^{-2}$, and $r=10^{3}, 1.2\times 10^{3}$, $c_{s}^{2}$ also decreases with an increase in $b$. The effect of increasing $r$, or equivalently $N$, is minimal.
\par
In Fig. \ref{fig12}, we plot the $b$ dependence of the parallel and perpendicular pressures, defined as $P_{\|}=P-B^{2}/2$ and $P_{\perp}=P-MB+B^{2}/2+v^{2}\epsilon/2$. These are examined for $R\Omega=0,0.5$ and at fixed $t=0.8$ in the QGP case [Fig. \ref{fig12}(a)] and $R\Omega=0,0.02$ and at $t=0.01$ in the NS case [Fig. \ref{fig12}(b)].
When $R\Omega=0$, the dependence of $P_{\|}$ and $P_{\perp}$ on $b$ resembles the dependence observed in quark models that discuss the chiral phase transition in the absence of rotation [see, e.g. \cite{fayazbakhsh2014}]. The effects of relativistic temperature corrections and the centrifugal term $v^{2}\epsilon/2$ in \eqref{C28} are negligible. Similar to the case when $\Omega=0$, $P_{\|}$ remains positive and increases with rising $b$, while $P_{\perp}$ is negative and decreases as $b$ increases.
\par
In addition to $P_{\|}$ and $P_{\perp}$, Fig. \ref{fig13} illustrates the $t$ dependence of the anisotropy factor $\xi_{\text{aniso}}$, defined in \eqref{C29}, for both the QGP [Fig. \ref{fig13}(a)] and the NS [Fig. \ref{fig13}(b)] scenarios. We observe that $\zeta_{\text{aniso}}$ decreases with increasing temperature. In the case of QGP, at a constant temperature, $\zeta_{\text{aniso}}$ increases as $v$ increases. However, for the NS case, the effects of a nonzero $v$ do not significantly alter the temperature dependence of $\xi_{\text{aniso}}$.
\section{Conclusions}\label{sec6}
\setcounter{equation}{0}
In this work, we investigated the BEC phenomenon in a charged Bose gas that is simultaneously subjected to a constant magnetic field and rigid rotation. Using the Fock-Schwinger formalism, we derived the free propagator for this model at finite temperature and density. This propagator was then used to compute the thermodynamic potential of the system. We focused on the nonrelativistic limit and the contribution of particles to this potential in the LLL approximation. To achieve a consistent thermodynamic potential, we appropriately modify the effective chemical potential. This process naturally led to the definition of a magnetorotational fugacity, $z_{B,\Omega}$. This quantity represents one of the central theoretical results of our work, providing a unified description of the combined effect of magnetic fields and rotation on the thermodynamic properties of the system. Within the high-temperature approximation used in this work, rigid rotation enters the thermodynamic description solely through the Tolman-Ehrenfest local temperature. Consequently, the rotational dependence of $z_{B,\Omega}$ is entirely encoded in the Tolman-Ehrenfest local temperature rather than appearing explicitly through the angular velocity. The Tolman correction accounts for the variation in temperature based on the distance $R$ from the rotation axis for a fixed angular velocity.
\par
The main goal of this work was to determine whether rigid rotation qualitatively affects the weak BEC scenario induced by a strong magnetic field. Our analysis reveals that the weak scenario, primarily driven by the Landau quantization and the resulting dimensional reduction, remains remarkably robust even in the presence of rigid rotation. While the Tolman correction quantitatively modifies the local thermodynamic quantities through the magnetorotational fugacity, it does not alter the condensation mechanism. Three independent observations consistently demonstrate the persistence of the weak scenario. Firstly, we showed that in the phenomenologically relevant temperature interval the magnetorotational fugacity is below one, preventing the formation of conventional Bose-Einstein condensates in these regimes. Secondly, by separating the particle number density into contributions from ground- and excited-states, and introducing a finite momentum cutoff $p_{0}$ around the lowest energy state, we explicitly constructed the finite neighborhood of the ground state that characterizes the weak BEC scenario. The ground state number density decreases as temperature increases. Finally, the specific heat remains smooth and exhibits no singularity. These results establish a self-consistent set of thermodynamic signatures that demonstrate the diffuse nature of weak BEC in the presence of magnetic fields and rigid rotation.
\par
Although rotation leaves the weak BEC mechanism intact, it produces several measurable thermodynamic signatures. Other thermodynamic quantities, such as pressure, angular-momentum and entropy densities, compressibility, susceptibilities, and moment of inertia, all receive rotational corrections through the Tolman temperature, which depends strongly on the characteristic scales of the system. These corrections become particularly significant under QGP conditions, but are strongly suppressed for NS parameters because of the much smaller linear velocities. We introduced a characteristic temperature $t_{3/4}$, which provides a quantitative measure of the gradual formation of the low-momentum population. Besides confirming the persistence of the weak BEC scenario, our analysis reveals a qualitatively new magnetic response of rotating charged Bose matter through a temperature-driven diamagnetic-to-paramagnetic crossover in the QGP regime.
The crossover temperature is controlled by both the magnetic-field strength and the rotational velocity. Increasing rotation enhances the paramagnetic response, while stronger magnetic fields reinforce diamagnetism. This phenomenon reflects the competition between Landau quantization and the enhancement of orbital angular-momentum due to rotation. In contrast, the NS regime remains diamagnetic across the relevant temperature range. The observable consequences of rotation are therefore controlled by the hierarchy of the characteristic physical scales rather than by changes in the underlying condensation mechanism.
\par
The present analysis has been performed for a noninteracting charged Bose gas within the nonrelativistic, lowest Landau level, and high-temperature approximations. Within these controlled approximations, the robustness of the weak BEC scenario has been established. To provide a more realistic setup and, in particular, to clarify the validity of these approximations, future studies could extend the present framework to include higher Landau levels, self-interactions, and finite-size effects. An important next step is to extend this framework to charged vector mesons, where intrinsic spin would provide a natural setting for exploring the interplay between magnetic fields, rotation, and spin in hot hadronic matter. The observed competition between rotation and magnetic fields, which is manifested through the diamagnetism-to-paramagnetism transition in this work, suggests that the interplay between magnetic fields and rigid rotation may significantly influence the angular-momentum response of hot QCD matter. Therefore, expanding the current formalism to include fermionic systems is expected to provide new insights into the origin and evolution of spin polarization observed in relativistic HICs.
\begin{appendix}
\section{Useful relations}\label{appA}
\setcounter{equation}{0}
To derive the thermodynamic relations in Sec. \ref{sec4}, we have used following relations:
\begin{eqnarray}\label{appB1}
\frac{\partial\lambda_{B}^{-1}}{\partial T}&=&\frac{1}{2T\lambda_{B}}, \qquad \frac{\partial \lambda_{B}^{-1}}{\partial (eB_{0})}=\frac{1}{4\lambda_{B}m_{B}^{2}},\nonumber\\
\frac{\partial m_{B}}{\partial (eB_{0})}&=&\frac{1}{2m_{B}},\qquad
\left(\frac{\partial z_{B,\Omega}}{\partial T}\right)_{z_{B}}=\frac{N\Omega}{T^{2}}z_{B,\Omega}, \nonumber\\
\left(\frac{\partial z_{B,\Omega}}{\partial T}\right)_{\mu,\Omega,B}&=&-\frac{\left(\mu_{B}-N\Omega\right)}{T^{2}}z_{B,\Omega}, \nonumber\\
\left(\frac{\partial z_{B,\Omega}}{\partial \mu}\right)_{\Omega,T,B}&=&\beta z_{B,\Omega},\nonumber\\
\left(\frac{\partial z_{B,\Omega}}{\partial \Omega}\right)_{\mu,T,B,v}&=&-\beta Nz_{B,\Omega},\nonumber\\
\left(\frac{\partial z_{B,\Omega}}{\partial (eB_{0})}\right)_{\mu,T,\Omega}&=&-\frac{\beta z_{B,\Omega}}{2m_{B}}.
\end{eqnarray}
Here, we used $z_{B,\Omega}=z_{B}e^{-\beta N\Omega}$ with $z_{B}=e^{\beta\mu_{B}}$ with $m_{B}^{2}=m^{2}+eB_{0}$.
\end{appendix}

\end{document}